\documentclass[twocolumn]{aastex63}
\pdfoutput=1 
\usepackage[T1]{fontenc}
\usepackage{amsmath,amstext}
\usepackage{apjfonts} 
\usepackage[figure,figure*]{hypcap}
\usepackage{microtype}
\usepackage{tablefootnote}
\newcommand{\repeater}{FRB~121102} 

\begin{document}
\shortauthors{Eftekhari et al.}
\shorttitle{Late-Time Radio and Millimeter Observations of SLSNe and LGRBs}

\uppercase{\title{\normalfont Late-Time Radio and Millimeter Observations of Superluminous Supernovae and Long Gamma Ray Bursts: Implications for Central Engines, Fast Radio Bursts, and Obscured Star Formation}}

\author[0000-0003-0307-9984]{T.~Eftekhari}
\affiliation{Center for Astrophysics | Harvard \& Smithsonian, Cambridge, MA 02138, USA}

\author[0000-0001-8405-2649]{B.~Margalit}
\altaffiliation{NASA Einstein Fellow}
\affiliation{Astronomy Department and Theoretical Astrophysics Center, University of California, Berkeley, Berkeley, CA 94720, USA}

\author[0000-0002-9646-8710]{C.~M.~B.~Omand}
\affiliation{Department of Physics, School of Science, The University of Tokyo, Tokyo 113-0033, Japan}

\author[0000-0002-9392-9681]{E.~Berger}
\affiliation{Center for Astrophysics | Harvard \& Smithsonian, Cambridge, MA 02138, USA}

\author[0000-0003-0526-2248]{P.~K.~Blanchard}
\affiliation{Center for Interdisciplinary Exploration and Research in Astrophysics (CIERA) and Department of Physics and Astronomy, Northwestern University, 1800 Sherman Ave, Evanston, IL 60201, USA}

\author{P.~Demorest}
\affiliation{National Radio Astronomy Observatory, Socorro, NM 87801, USA}

\author[0000-0002-4670-7509]{B.~D.~Metzger}
\affiliation{Department of Physics and Columbia Astrophysics Laboratory, Columbia University, New York, NY 10027, USA}

\author[0000-0002-5358-5642]{K.~Murase}
\affiliation{Department of Physics, The Pennsylvania State University, University Park, PA 16802, USA}
\affiliation{Department of Astronomy \& Astrophysics, The Pennsylvania State University, University Park, PA 16802, USA}
\affiliation{Center for Multimessenger Astrophysics, The Pennsylvania State University, University Park, PA 16802, USA}
\affiliation{Yukawa Institute for Theoretical Physics, Kyoto University, Kyoto 606-8502, Japan}

\author[0000-0002-2555-3192]{M.~Nicholl}
\affiliation{Institute for Astronomy, University of Edinburgh, Royal Observatory, Blackford Hill, Edinburgh EH9 3HJ, UK}
\affiliation{Birmingham Institute for Gravitational Wave Astronomy and School of Physics and Astronomy, University of Birmingham, Birmingham B15 2TT, UK}

\author[0000-0002-5814-4061]{V.~A.~Villar}
\affiliation{Center for Astrophysics | Harvard \& Smithsonian, Cambridge, MA 02138, USA}

\author[0000-0003-3734-3587]{P.~K.~G.~Williams}
\affiliation{Center for Astrophysics | Harvard \& Smithsonian, Cambridge, MA 02138, USA}
\affiliation{American Astronomical Society, Washington, DC 20006, USA}

\author[0000-0002-8297-2473]{K.~D.~Alexander}
\altaffiliation{NASA Einstein Fellow}
\affiliation{Center for Interdisciplinary Exploration and Research in Astrophysics (CIERA) and Department of Physics and Astronomy, Northwestern University, 1800 Sherman Ave, Evanston, IL 60201, USA}

\author[0000-0002-2878-1502]{S.~Chatterjee}
\affiliation{Cornell Center for Astrophysics and Planetary Science and Department of Astronomy, Cornell University, Ithaca, NY 14853, USA}

\author[0000-0001-5126-6237]{D.~L.~Coppejans}
\affiliation{Center for Interdisciplinary Exploration and Research in Astrophysics (CIERA) and Department of Physics and Astronomy, Northwestern University, 1800 Sherman Ave, Evanston, IL 60201, USA}

\author[0000-0002-4049-1882]{J.~M.~Cordes}
\affiliation{Cornell Center for Astrophysics and Planetary Science and Department of Astronomy, Cornell University, Ithaca, NY 14853, USA}

\author[0000-0001-6395-6702]{S.~Gomez}
\affiliation{Center for Astrophysics | Harvard \& Smithsonian, Cambridge, MA 02138, USA}

\author[0000-0002-0832-2974]{G.~Hosseinzadeh}
\affiliation{Center for Astrophysics | Harvard \& Smithsonian, Cambridge, MA 02138, USA}

\author{B.~Hsu}
\affiliation{Center for Astrophysics | Harvard \& Smithsonian, Cambridge, MA 02138, USA}

\author[0000-0003-4299-8799]{K.~Kashiyama}
\affiliation{Department of Physics, School of Science, The University of Tokyo, Tokyo 113-0033, Japan}
\affiliation{Research Center for the Early Universe, The University of Tokyo, Tokyo 113-0033, Japan}

\author[0000-0002-8297-2473]{R.~Margutti}
\affiliation{Center for Interdisciplinary Exploration and Research in Astrophysics (CIERA) and Department of Physics and Astronomy, Northwestern University, 1800 Sherman Ave, Evanston, IL 60201, USA}

\author[0000-0002-5723-8023]{Y.~Yin}
\affiliation{Center for Astrophysics | Harvard \& Smithsonian, Cambridge, MA 02138, USA}

\begin{abstract}
We present the largest and deepest late-time radio and millimeter survey to date of superluminous supernovae (SLSNe) and long duration gamma-ray bursts (LGRBs) to search for associated non-thermal synchrotron emission. Using the Karl G. Jansky Very Large Array (VLA) and the Atacama Large Millimeter/submillimeter Array (ALMA), we observed 43 sources at 6 and 100 GHz on a timescale of $\sim 1 - 19$ yr post-explosion. We do not detect radio/mm emission from any of the sources, with the exception of a 6 GHz detection of PTF10hgi \citep{Eftekhari2019}, as well as the detection of 6 GHz emission near the location of the SLSN PTF12dam, which we associate with its host galaxy. We use our data to place constraints on central engine emission due to magnetar wind nebulae and off-axis relativistic jets. We also explore non-relativistic emission from the SN ejecta, and place constraints on obscured star formation in the host galaxies. In addition, we conduct a search for fast radio bursts (FRBs) from some of the sources using VLA Phased-Array observations; no FRBs are detected to a limit of $16$ mJy ($7\sigma$; 10 ms duration) in about 40 min on source per event. A comparison to theoretical models suggests that continued radio monitoring may lead to detections of persistent radio emission on timescales of $\gtrsim {\rm decade}$.
\end{abstract}


\section{Introduction}\label{sec:intro}

The advent of wide-field untargeted optical surveys has led to the discovery of hydrogen-poor superluminous supernovae (SLSNe\footnote{For simplicity we refer to the hydrogen poor (Type I) events as SLSNe.  The hydrogen-rich events appear to simply be an extension of the Type IIn SN population.}), a rare class of core-collapse supernovae (SNe) with luminosities up to 100 times larger than ordinary SNe (e.g., \citealt{Chomiuk2011,Quimby2011,Gal-Yam2012}). The energy source of SLSNe has been a topic of debate, with ideas ranging from pair-instability explosions \citep{Gal-Yam2009} to interaction with a dense hydrogen-poor circumstellar medium \citep{Chevalier2011}, and the spin-down of a millisecond magnetar central engine \citep{Kasen2010,Woosley2010}. In recent years, a growing line of evidence has emerged in favor of a central engine origin. Indeed, the light curve evolution of these events can be well-characterized by the dipole spin-down of strongly magnetized neutron stars with initial spin periods of $\sim 1 - 10$ ms and large magnetic fields of $\sim 10^{13} - 10^{15}$ G \citep{Inserra2013,Nicholl2014,Nicholl2017b,Blanchard2020}.

Some similarities have also been noted between SLSNe and long duration gamma-ray bursts (LGRBs).  Namely, both are rare explosions that arise from stripped massive stars, exhibit a preference for low metallicity host galaxies (e.g., \citealt{Modjaz2008,Levesque2010,Lunnan2014}), share ejecta properties as evidenced by nebular-phase spectra  \citep{Nicholl2016b,Jerkstrand2017,Nicholl2019}, and appear to be powered by central engines \citep{Mazzali2014,Margalit2018a}, although in the case of LGRBs black hole engines have also been assumed (e.g., \citealt{MacFadyen1999,Metzger2011}).

Another intriguing connection to SLSNe has been suggested by the discovery and localization of the repeating fast radio burst \repeater{} \citep{Spitler2014,Spitler2016} to a low metallicity star forming galaxy \citep{Chatterjee2017,Tendulkar2017} and an associated parsec-scale persistent radio source \citep{Marcote2017}, whose large and variable rotation measure indicates a highly magnetized and dynamic environment \citep{Michilli2018}. These properties have prompted theories suggesting that FRB production is associated with the birth of young, millisecond magnetars in SLSN and/or LGRB explosions \citep{Murase2016,Piro2016,Metzger2017,Kashiyama2017,Nicholl2017b,Margalit2018a,Margalit2018b}.

More recently, the localizations of 10 apparently non-repeating FRBs \citep{Bannister2019,Prochaska2019,Ravi2019,Bhandari2020,Macquart2020,Law2020,Heintz2020} to more massive galaxies suggests that some FRBs may be produced by magnetars formed from an older progenitor population such as binary neutron star (BNS) mergers or accretion-induced collapse (AIC) of white dwarfs \citep{Margalit2019}. Indeed, the recent discovery of a luminous, millisecond-duration radio burst from the Galactic magnetar SGR 1935+2154 \citep{Scholz2020,Bochenek2020} may support a connection between magnetars and FRBs \citep{Margalit2020}. Finally,  two additional repeating FRBs (FRBs 180916 and 190711) have been localized to host galaxies \citep{Marcote2020,Heintz2020}. Similar to the host of \repeater{}, the host galaxies of these events are both less luminous and less massive than the host galaxies of the apparently non-repeating FRBs, though they exhibit a range of star formation rates \citep{Heintz2020}. 

The long term radio and millimeter properties of SLSNe, which may shed light on their energy source and probe a connection with \repeater-like systems, have not been explored to date. Early radio follow up (timescales of $\lesssim {\rm 1 - 8 \ yr}$) has yielded only non-detections (e.g., \citealt{Coppejans2018,Hatsukade2018}).  On the other hand, follow-up of the SLSN PTF10hgi on a timescale of about $8-10$ years post-explosion has led to a number of radio detections \citep{Eftekhari2019,Law2019,Mondal2020}. While a magnetar origin has not yet been confirmed for this source, the available data are consistent with emission powered by a central engine, either from a magnetar-powered nebula or an off-axis jet \citep{Eftekhari2019}. 

In addition to searches for radio nebulae from putative central engines, radio and mm observations provide insight into the properties of non- or mildly-relativistic ejecta (e.g., \citealt{Berger2002}), as well as obscured star formation in the host galaxies of these events (e.g., \citealt{Berger2003a,Hatsukade2018,Greiner2016}).

Here we present deep late-time radio and mm observations of 43 SLSNe and LGRBs to search for non-thermal synchrotron emission from these sources, which may shed light on their central engines, connection to FRBs, slow ejecta, and host galaxies. Our study provides the largest and deepest sample of radio and mm observations of SLSNe to date. We use the data to explore  magnetar wind nebulae, off-axis jets, emission from the non-relativistic SN ejecta, and obscured star formation in the host galaxies. The paper is structured as follows. We present our observations in \S\ref{sec:obs}. Constraints on obscured star formation in the host galaxies are presented in \S\ref{sec:sfr}. In \S\ref{sec:snejecta} and \S\ref{sec:jets}, we examine synchrotron emission due to non-relativistic and relativistic outflows, respectively. We place constraints on putative magnetar wind nebulae in \S\ref{sec:nebula}, and we summarize our results in \S\ref{sec:conc}. Throughout the paper, we use the latest Planck cosmological parameters, $H_0 = \rm 67.8 \ km \ s^{-1} \ Mpc^{-1}$, $\Omega_m = 0.308$, and $\Omega_\Lambda = 0.692$ \citep{Planck2016}.

\section{VLA and ALMA Observations}
\label{sec:obs}

\subsection{Sample Selection}\label{sec:sources}

The SLSNe observed with the Karl G.~Jansky Very Large Array (VLA) represent a complete (as of February 2017) volume-limited sample to a distance twice that of FRB\,121102 ($z\lesssim 0.35$), and with a timescale of $\gtrsim 5$ years post-explosion, leading to 15 SLSNe. Of these, 3 events were previously observed in the radio at early time, leading to non-detections: PTF09cnd \citep{Chandra2009,Chandra2010}, SN2012il \citep{Chomiuk2012}, and SN2015bn \citep{Nicholl2016a,Nicholl2018}; see Table~\ref{tab:litslsne}. Observations of PTF10hgi obtained as part of this work were presented in \citet{Eftekhari2019}. We also observed seven LGRBs with the VLA, of which two were detected at early times: (i) GRB\,020903 had a 5 GHz flux density of $\approx 100$ $\mu$Jy at 0.5 year post-explosion, declining as $F \propto t^{-1.1}$ \citep{Soderberg2004}; and (ii) GRB\,030329 was last detected 4.9 years post-explosion with a 5 GHz flux density of $70$ $\mu$Jy and a decline rate of $t^{-1.3}$ \citep{Mesler2012}. The contribution from an afterglow at the time of our observations ($\delta t\approx 5490$ and $\delta t\approx 5323$ d for GRB\,020903 and GRB\,030329, respectively) is expected to be negligible. 

We additionally observed 29 SLSNe with the Atacama Large Millimeter/submillimeter Array (ALMA), seven of which overlap with our VLA sample. These events were observed as part of two observing campaigns. Events observed during the first campaign (2017.1.00280.S; PI: Berger) constitute all sources accessible to ALMA (as of March 2017)\footnote{The sole exception is PS1-14bj, which was not observed due to a scheduling error.} with a timescale of $\gtrsim 3$ years post-explosion and a distance of $z\lesssim 0.5$. The second campaign (2019.1.01663.S; PI: Eftekhari) consists of all events with Decl.~$<30^\circ$ and $z<0.4$, with the exception of SN2017egm which has been observed as part of a dedicated ALMA campaign; the source is also observed at early times in the radio \citep{Bright2017,Bose2018,Coppejans2018}.

\subsection{VLA Continuum Observations}

We obtained 6 GHz (C-band) radio observations of our sample with the VLA in configurations A and B with a total on-source integration time of $\approx 40$ min per target. The details of the observations are summarized in Table~\ref{tab:vla}. All observations obtained in 2017 utilized the 8-bit samplers with $\sim 2$ GHz bandwidth (excluding excision of edge channels and RFI), while our 2019 observations were configured using the 3-bit samplers, providing the full 4 GHz of bandwidth across the observing band. 

We processed the data within the Common Astronomy Software Application (CASA) software package \citep{McMullin2007}. We performed bandpass and flux density calibration using 3C286 and 3C48. Complex gain calibrators for individual sources are listed in Table~\ref{tab:vla}. We imaged each field using $2048\times 2048$ pixels at a scale of $0.07-0.2$ arcsec per pixel using multi-frequency synthesis (MFS; \citealt{Sault1994}) and $w$-projection with 128 planes \citep{Cornwell2008}. Flux densities at the source positions (and image rms values) were extracted using the \texttt{imtool} program as part of the \texttt{pwkit}\footnote{Available at https://github.com/pkgw/pwkit.} package \citep{Williams2017} (Table~\ref{tab:vla}).
\begin{figure}
\vspace{0.1cm}
\includegraphics[width=\columnwidth]{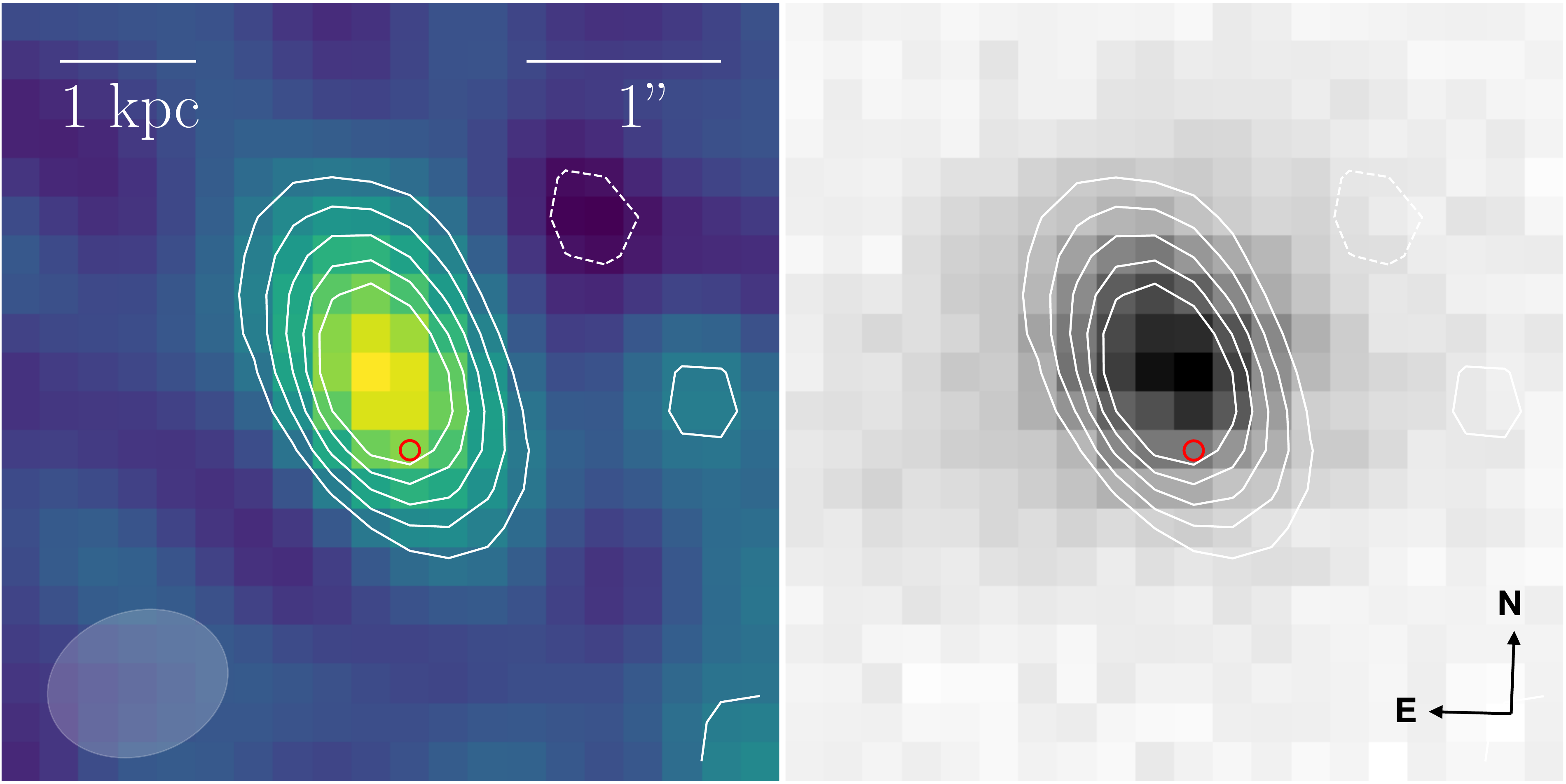}
\caption{\textit{Left:} Radio continuum image of the host galaxy of PTF12dam from our VLA 6 GHz (C band) observation in B-configuration. Contours correspond to $-2$, 2, 3, 4, 5, and 6 times the rms noise of the image. The synthesized beam (0.95'' $\times$ 0.74'') is shown in the lower left corner. Shown as a red circle is the optical position of the SN. \textit{Right:} PanSTARRS optical $i$-band image of the host galaxy with radio contours overlaid.}
\label{fig:ptf12dam}
\end{figure}
In addition to the detection of PTF10hgi, with $F_\nu = 47 \pm 7 \ \mu$Jy, reported in \citet{Eftekhari2019}, we detect radio emission near the location of the SLSN PTF12dam with a flux density $F_\nu = 117\pm 12 \ \mu$Jy; see Figure~\ref{fig:ptf12dam}. This is comparable to the 3 GHz detection presented in \citet{Hatsukade2018} with $F_\nu = 141.5 \pm 5.1 \ \mu$Jy.
The radio emission is offset from the SN position by $\sim 1$ arcsecond and is instead centered on, and traces, the optical emission from the host galaxy. Thus the emission is most likely related to star formation in the host galaxy (see discussion in \S\ref{sec:sfr}). Indeed, the emission is resolved out in our second epoch, A-configuration observation with a decreased flux density $F_\nu = 61.5 \pm 10.8 \ \mu$Jy, indicative of an extended source.  No clear radio emission (SNR $> 5 \sigma$) is detected from the remainder of the SLSNe in our sample, or from the location of any of the LGRBs in our sample, with typical rms values of $5-10$ $\mu$Jy. 

We report marginal detections near the positions of SN2009jh and GRB\,050826. In the case of SN2009jh, the source significance is $3.6\sigma$ ($F_\nu = 23.3\pm 6.4 \ \mu$Jy) but with an offset of $1.6\pm 0.3''$ from the SN position, and we therefore consider it unlikely to be related to the SN.  Similarly, for GRB\,050826, the source significance is only $2.5\sigma$ ($F_\nu = 23\pm 9 \ \mu$Jy) with an offset of $1.0\pm 0.5''$ from the GRB position.

\subsection{Phased VLA Observations}

In addition to standard continuum observations, we also obtained phased-array observations with the VLA to search for FRBs from the observed sources. Similar searches for FRBs from the locations of SLSNe and LGRBs were presented in \citet{Law2019} and \citet{Hilmarsson2020}, yielding no detections. We present here the results of our initial VLA sample of events observed in 2017. The phased-array data were recorded with 256 $\mu$s time resolution and 2 MHz channels with 2 GHz total bandwidth. Each raw filterbank file is divided into a channelized time series with 1 GHz bandwidth centered at 5 and 7 GHz. We performed a standard RFI search using the \texttt{rfifind} routine in \texttt{PRESTO} \citep{Ransom2001} with two second integration times. The RFI-excised maps are applied to the data for subsequent processing. For each source, we incoherently dedispersed the data at $1000$ trial DMs ranging up to $\rm DM = 5000 \ \rm pc \ cm^{-3}$ with a step size of $5$, corresponding to a maximum redshift $z\sim 7$, well beyond the max redshift of our sample ($z\sim 0.57$), and indeed well above the maximum observed DM for any FRB observed to date\footnote{http://frbcat.org/}. This additionally allows for a substantial host galaxy DM contribution. Following dedispersion, we normalize the time series by performing a standard red noise removal. Individual scans are then searched for FRBs using the matched-filtering algorithm \texttt{single\_pulse\_search.py} \citep{Ransom2001}.

Following the prescription of \citet{Cordes2003}, the minimum detectable flux density for an FRB is given by:
\begin{equation}
S_{\rm min} = \dfrac{\rm (S/N)_{\rm min}SEFD}{\sqrt{n_{\rm pol} \Delta \nu W}}
\end{equation}
where $n_{\rm pol}$ is the number of summed polarizations, $\Delta \nu$ is the bandwidth, $W$ is the intrinsic pulse width, $(S/N)_{\rm min}$ is the minimum signal-to-noise threshold, and SEFD refers to the system equivalent flux density. Assuming a phasing efficiency factor of $0.9$ and a nominal $10$ ms pulse width, we find a minimum detectable flux density of $S_{\rm min}\approx 16$ mJy ($7\sigma$) for our observations. 

We do not detect any FRBs from our sample. We place limits on the maximum energy of an FRB emitted from each source following the prescription of \citet{Cordes2019}, in which the burst energy is given by:
\begin{equation}
E_{\rm b,max} = 4\pi S_{\nu} W \Delta \nu d_L f_b, 
\end{equation}
where we assume a beaming fraction $f_b = 0.5$. For the range of redshifts probed by our initial VLA sample of SLSNe ($z \sim 0.101 - 0.376$), we find $E_{\rm b,max} \lesssim 2 \times 10^{37} - 4 \times 10^{38}$ erg. This is comparable to the range of inferred burst energies for the lowest energy bursts from \repeater{}, with $E_b \sim 10^{37} - 10^{38}$ erg \citep{Gourdji2019}. Thus, while we are sensitive to the low energy range of FRBs (from \repeater), given the intermittent nature of \repeater, the short duration of our observations ($\approx 40$ min per source) is prohibitively small. A significant time investment with the GBT, Arecibo, or FAST may reveal bursts from these sources.

\subsection{ALMA Observations}

We obtained millimeter observations with ALMA in Band 3 ($\sim$ 100 GHz) of 28 SLSNe. Flux density and complex gain calibrators are listed in Table~\ref{tab:alma}. In all cases, we use the ALMA data products, which utilize standard imaging techniques within CASA. Each field is imaged using MFS, Briggs weighting with a robust parameter of 0.5, and a standard gridding convolution function. Image scales span $0.04-0.2$ arcseconds per pixel, with typical image sizes of about $1500\times 1500$ pixels. Flux densities and image rms values were obtained using the \texttt{imtool} program as part of the \texttt{pwkit} package.  

We do not detect millimeter emission from any of our sources with the exception of a $3\sigma$ detection near the location of SN2007bi, with a flux density $F_\nu = 55.3 \pm 18.8 \ \mu$Jy  (R.A.=${\rm 13^h19^m20^s.226}$, decl.=$+08^\circ55'43''.85$; J2000). To compare the position of the millimeter source relative to the SN position,  we use archival images of SN2007bi from the Liverpool Telescope \citep{Young2010} which are first matched to an absolute astrometric system. We find that the millimeter source is offset from the SN position by $1\pm 0.6''$ and is thus unlikely to be related to the SN. We similarly compare the position of this source to that of the host galaxy of SN2007bi using a wide-field $i$-band image from the Panoramic Survey Telescope and Rapid Response System (Pan-STARRS1) $3\pi$ survey. We first perform relative astrometry between the PS1/$3\pi$ image and the Liverpool image and find an astrometric tie uncertainty of $\sigma_{\rm host}\approx 0.6''$. Comparing the position of the millimeter source to this image, we find that the source is offset from the host galaxy position by $1.5\pm 0.6''$. We thus conclude that the weak ALMA detection is unlikely to be related to the SN or the host galaxy. 

\begin{figure}
\includegraphics[width=1.05\columnwidth]{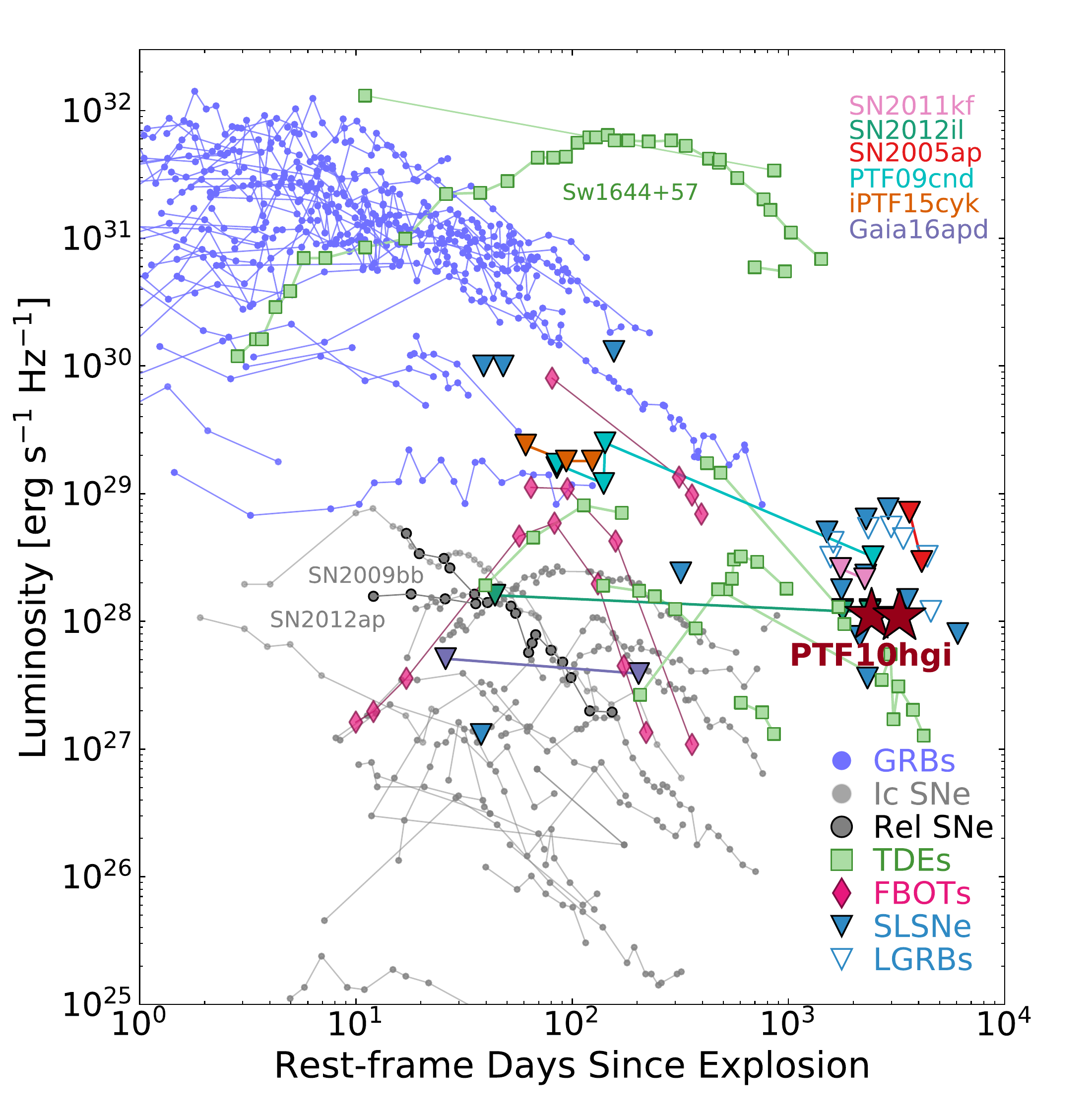}
\caption{Radio luminosities at 6 GHz for our sample of SLSNe (blue filled triangles) and LGRBs (blue open triangles), compared to the 6$-$10 GHz luminosities of other classes of radio transients, including relativistic SNe \citep{Soderberg2010,Margutti2014}, ``normal'' hydrogen-poor core-collapse Ibc SNe \citep{Soderberg2005,Soderberg2006,Chomiuk2012}, fast blue optical transients (\citealt{Margutti2019,Coppejans2020,Ho2019,Ho2020}; Coppejans et al. \textit{in prep.}), GRBs \citep{Chandra2012}, and tidal disruption events \citep{Zauderer2011,Berger2012,Cenko2012,Zauderer2013,Irwin2015,Alexander2017_tde,Brown2017,Saxton2017,Eftekhari2018,Mattila2018}. We highlight SLSNe with multiple epochs of observations from the literature (references are given in Table~\ref{tab:litslsne}).}
\label{fig:lums}
\end{figure}

\subsection{Comparison to Other Transients}
\label{sec:radio_trans}

In Figure~\ref{fig:lums} we plot the 6 GHz radio luminosity upper limits of our sample of SLSNe and LGRBs, compared to the $6-10$ GHz light curves of other classes of transients, including relativistic SNe, normal Type Ib/c SNe, fast blue optical transients (FBOTs), and tidal disruption events (TDEs). We also include observations of LGRBs and SLSNe from the literature. 

We find that the late-time limits presented here probe a largely unexplored region of parameter space with timescales of $\delta t \gtrsim 10^3$ d and luminosities $L_{\nu} \sim 10^{28} - 10^{29} \ \rm erg \ s^{-1} \ Hz^{-1}$. With the exception of a few TDEs, most transients have been observed in the radio at earlier times. The radio limits for SLSNe are significantly fainter than the vast majority of cosmological GRBs, which are observed at earlier times ($\delta t \sim 1 - 10^3$ d) with luminosities $L_{\nu} \gtrsim 10^{30} \ \rm erg \ s^{-1} \ Hz^{-1}$. Conversely, the limits do not reach the levels of the least luminous Type Ib/c SNe with $L_{\nu} \sim 10^{26} \ \rm erg \ s^{-1} \ Hz^{-1}$.   We explore the implications of these radio upper limits in the subsequent sections.

\section{Obscured Star Formation}
\label{sec:sfr}

Measurements of star formation rates as inferred from UV and optical data suffer from the obscuration effects of interstellar dust. At longer wavelengths, radio and millimeter observations offer a useful probe of star formation in dust-obscured galaxies as the emission arises from supernovae-accelerated cosmic ray electrons and dust heating \citep{Helou1985,Condon1991}. 
 
We place limits on the obscured SFRs in the host galaxies of our sample using the expression from \citet{Greiner2016}, which extrapolates the 1.4 GHz derived SFR of \citet{Murphy2011b} assuming a power law $F_\nu \propto \nu^{\alpha}$ and accounting for proper $k$-corrections:
\begin{equation}
{\rm SFR_{radio}} = 0.059\, {\rm M_{\odot} \ yr^{-1}}\, F_{\rm \nu,\mu Jy}d_{L,{\rm Gpc}}^2\nu_{\rm GHz}^{-\alpha}(1+z)^{-(\alpha + 1)},
\end{equation}
where $F_\nu$ is the observed flux density at a frequency $\nu$, and $d_L$ is the luminosity distance at a redshift $z$. We adopt $\alpha = 0.75$ for the spectral index \citep{Condon1992}. 

\begin{figure}
\includegraphics[width=1.1\columnwidth]{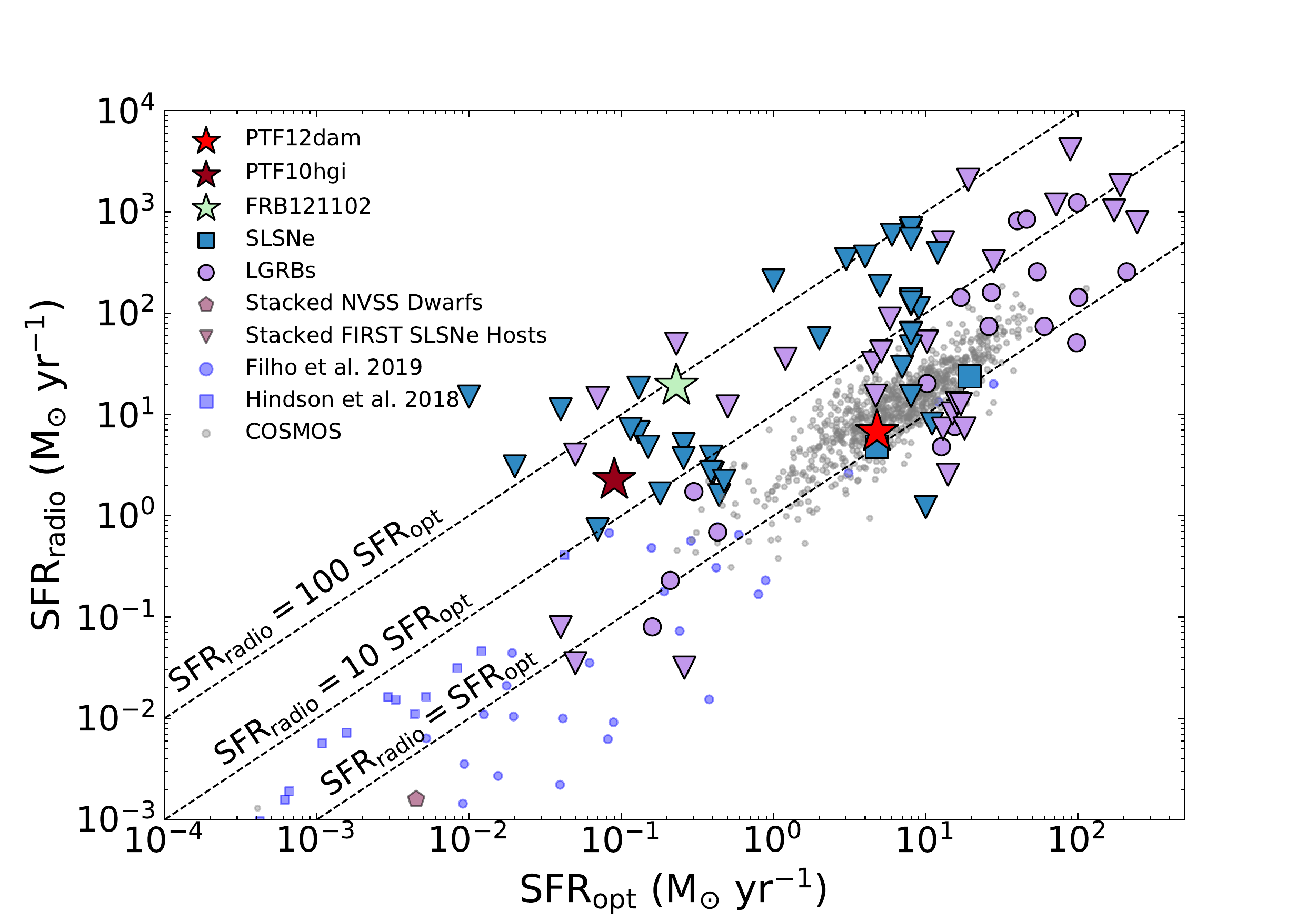}
\caption{Radio versus optical star formation rates for SLSNe (blue) and LGRBs (purple). Triangles correspond to $3\sigma$ upper limits. Also shown are our detections for PTF10hgi \citep{Eftekhari2019} and PTF12dam, as well as FRB 121102 (assuming a star formation origin for the radio emission). We also include radio observations of LGRB hosts (\citealt{Perley2013a}; \citealt{Greiner2016}; \citealt{Perley2015}, \citealt{Peters2019}) and SLSNe hosts \citep{Hatsukade2018} from the literature. Shown for comparison are nearby dwarf galaxies from a number of surveys \citep{Roychowdhury2012,Hindson2018,Filho2019}, as well as a sample of star forming galaxies at $z\lesssim 0.5$ from the VLA-COSMOS source catalog (gray dots; \citealt{Smoli2017}).  Dashed lines indicate  $\rm SFR_{radio} = SFR_{opt}, 10 \ SFR_{opt}$, and $\rm 100\ SFR_{opt}$.}
\label{fig:sfrs}
\end{figure}

In Figure~\ref{fig:sfrs}, we compare the radio/mm-inferred upper limits on SFR to the SFRs derived from H$\alpha$ measurements or SED modeling (references are given in Appendix~\ref{appendix:a}). We also plot radio observations of LGRB \citep{Perley2013a,Greiner2016,Perley2015} and SLSNe \citep{Hatsukade2018} hosts from the literature. For comparison, we include a sample of dwarf galaxies from a number of surveys \citep{Roychowdhury2012,Hindson2018,Filho2019}, as well as star forming galaxies at $z\lesssim 0.5$ from the VLA-COSMOS source catalog \citep{Smoli2017}. 

In the case of PTF12dam, our radio detection of the host galaxy corresponds to an SFR of $6.7\pm 0.7 \ \rm M_{\odot} \ yr^{-1}$, comparable to the value inferred from a previous 3 GHz detection ($4.8\pm 0.2 \ \rm M_{\odot} \ yr^{-1}$; \citealt{Hatsukade2018}) and an H$\alpha$ measurement, $4.8\pm 1 \ \rm M_{\odot} \ yr^{-1}$ \citep{Perley2016a}.  This indicates that the star formation in the host is largely unobscured. X-ray emission is also detected near the location of PTF12dam \citep{Margutti2018}, which we conclude is likely related to the host. Finally, we note that the host galaxy of PTF12dam is among the brightest SLSNe hosts with strong emission lines and an absolute magnitude $M_g = -19.33$ \citep{Chen2015}. 

Conversely, for PTF10hgi, \citet{Eftekhari2019} argue that the difference between the radio-inferred SFR of $2.3\pm 0.3 \ \rm M_{\odot} \ yr^{-1}$, and the optically-inferred SFR of $0.09 \ \rm M_{\odot} \ yr^{-1}$, indicates that the radio emission is not due to obscured star formation, but is instead related to the SN itself. The presence of weak emission lines from the host galaxy furthermore suggest that the system is not in an active starburst phase \citep{Perley2016a}. 

Finally, we note that the only remaining SLSN with a host galaxy radio detection is PTF10uhf \citep{Hatsukade2018}, in which the radio and optical SFR estimates are comparable, thus suggesting no dust obscuration as in the case of PTF12dam. Moreover, the host of PTF10uhf is unique relative to the population of SLSNe, with evidence of a merger between a large spiral galaxy and a less massive disk galaxy \citep{Perley2016a}. Indeed, when compared to the population of SLSNe host galaxies, the host of PTF10uhf is both among the most massive and the most prodigiously star-forming \citep{Perley2016a}.

For the remaining SLSNe in our sample, the radio-inferred SFR limits exceed the optically-inferred rates by factors of $\approx 2-5.2\times 10^3$ (Table~\ref{tab:master}), where the high end is driven primarily by the ALMA 100 GHz observations, which are less constraining in this context. The median of the ratio of radio-to-optical SFRs for the VLA sample is $<32$, indicating there is no evidence for significant dust obscuration in these SLSN host galaxies. 

We note that similar studies of SLSNe hosts in the radio are limited. Such studies include a search for radio emission from SLSNe host galaxies located within the footprints of wide-field radio surveys \citep{Schulze2018}, however the survey rms levels do not provide particularly meaningful constraints. Deeper VLA observations of eight additional SLSNe show no evidence for dust attenuation \citep{Hatsukade2018,Schulze2018}.

Among the LGRBs in our sample, three overlap with the radio sample presented in \citet{Michaowski2012}. The authors report 3$\sigma$ limits for the SFRs at 1.4 GHz using the prescription of \citet{Bell2003} which differs in normalization factor with our Equation 1. For consistency, we recompute their SFRs, finding $\rm SFR_{\rm radio} <6.5 \ \rm M_{\odot} \ yr^{-1}$ (GRB\,020903), $<21.0$ (GRB\,030329), and $<27.2$ (GRB\,061021). We improve on the limits for GRB\,030329 and GRB\,061021, finding $\rm SFR_{\rm radio} <2.5$ and $<13$, respectively. 

Limits on the radio-inferred SFRs for the host galaxies are roughly consistent with the optically-derived SFRs (within a factor of a few to ten, with the exception of GRB061021 where the radio limit is a factor of $\sim 260$ larger than the optical SFR), therefore ruling out significant dust obscuration in the hosts of the LGRBs for the majority of our sample.

To date, roughly 100 LGRB hosts have been observed in the radio, but the significant majority do not exhibit any discernible radio emission (e.g., \citealt{Berger2001,Michaowski2015,Perley2015,Greiner2016}). Radio-derived limits on the SFRs of these galaxies increasingly suggest little to no dust obscuration out to $z \approx 2$ (\citealt{Michaowski2012}; \citealt{Perley2015}; \citealt{Greiner2016}) (although a small subset of the population, referred to as ``dark'' GRBs, exhibit significant extinction \citep{Perley2013b}). Our limits are consistent with this scenario, allowing us to rule out significant dust obscuration for the majority of the LGRBs in our sample. We note that GRB\,090417B is considered a heavily-obscured, ``dark'' GRB \citep{Perley2013a}, where our radio-inferred SFR is a factor of $10$ times larger than the optical rate. 

\begin{figure}
\vspace{-0.3cm}
\includegraphics[width=1.1\columnwidth]{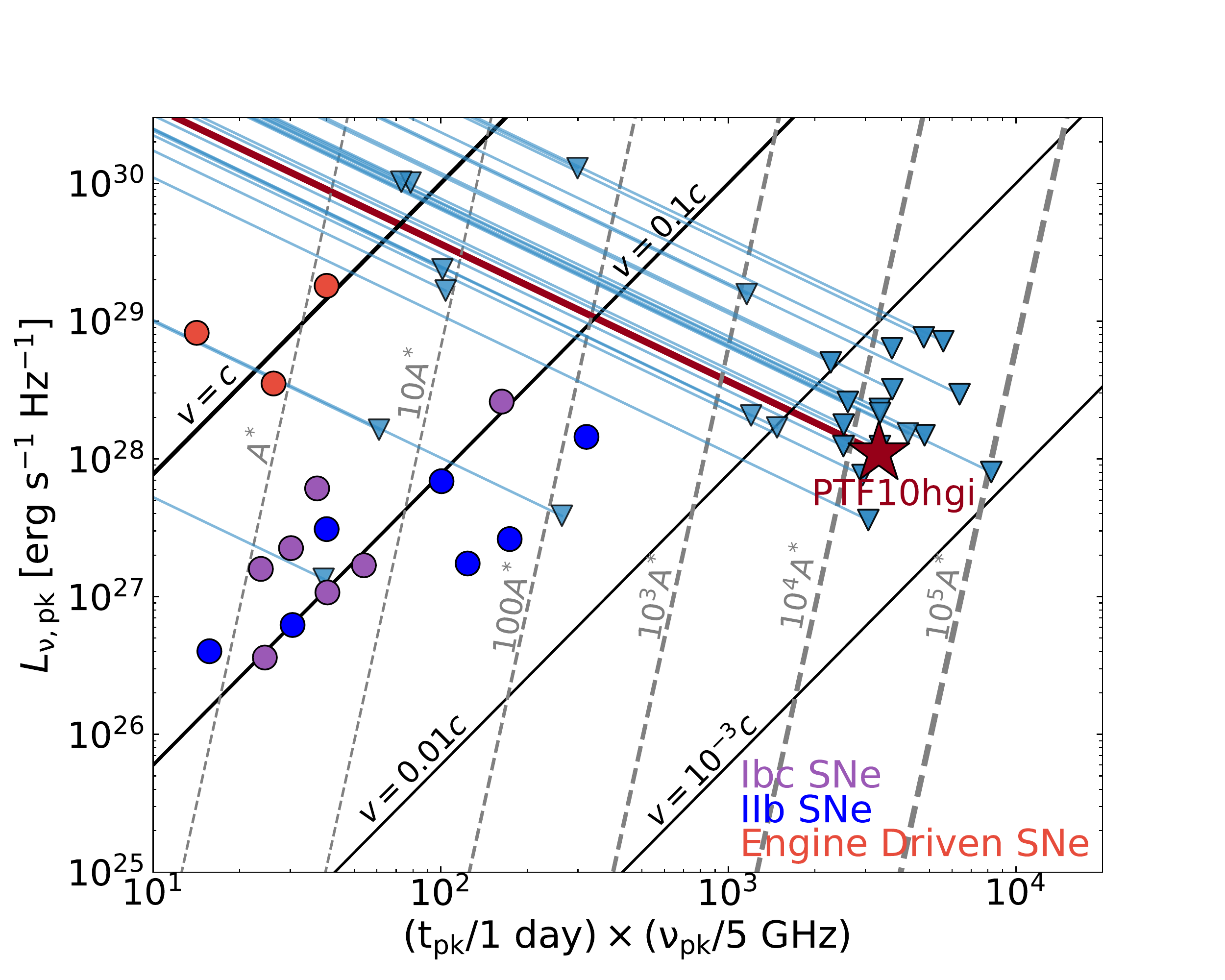}
\caption{Peak radio luminosity ($L_{\nu,\rm pk}$) versus the product of peak frequency and time ($\nu_{\rm pk} \times t_{\rm pk}$) for our sample of SLSNe (blue triangles), as well as all existing observations of SLSNe from the literature (see Table~\ref{tab:litslsne}). Individual lines for each SLSN account for a possible peak at an earlier time assuming $L_{\nu,\rm pk}\propto t^{-1}$. Black and grey lines mark constant shock velocity and mass-loss rate, respectively, given the standard formulation for synchrotron self-absorbed emission from a freely expanding non-relativistic blastwave \citep{Chevalier1998}, with $\epsilon_B = \epsilon_e = 0.1$. Also shown is the 6 GHz detection for PTF10hgi (star; \citealt{Eftekhari2019}) as well as Type Ib/c (purple circles), IIb (blue circles), and engine driven (red circles) SNe from the literature \citep{Soderberg2005,Margutti2019}.}
\label{fig:shock}
\end{figure}

Finally, we place tighter constraints on the mean radio and mm emission from each population (LGRBs and SLSNe) by stacking the individual images (excluding PTF12dam and PTF10hgi). We median-combine the images after centering on the transient position. We do not detect emission in either of the stacked populations, achieving 6 GHz rms values of about $2.6$ and $1.9$ $\mu$Jy for LGRBs and SLSNe, respectively. The stacked ALMA image yields an rms of $3.7 \ \mu$Jy. We translate the rms values into limits on the mean obscured SFR assuming the median redshift for each population. For the seven LGRBs in our sample, the non-detection corresponds to a $3\sigma$ limit of ${\rm SFR}\lesssim 4.2$ M$_{\odot}$ yr$^{-1}$ ($z\approx 0.3$). For the SLSNe, the $3\sigma$ SFR limits are $\lesssim 1.7$ and $\lesssim 27$ M$_{\odot}$ yr$^{-1}$ based on the 6 GHz ($z \approx 0.23$) and 100 GHz ($z \approx 0.23$) images, respectively.

\begin{figure}
\vspace{-0.3cm}
\includegraphics[width=1.1\columnwidth]{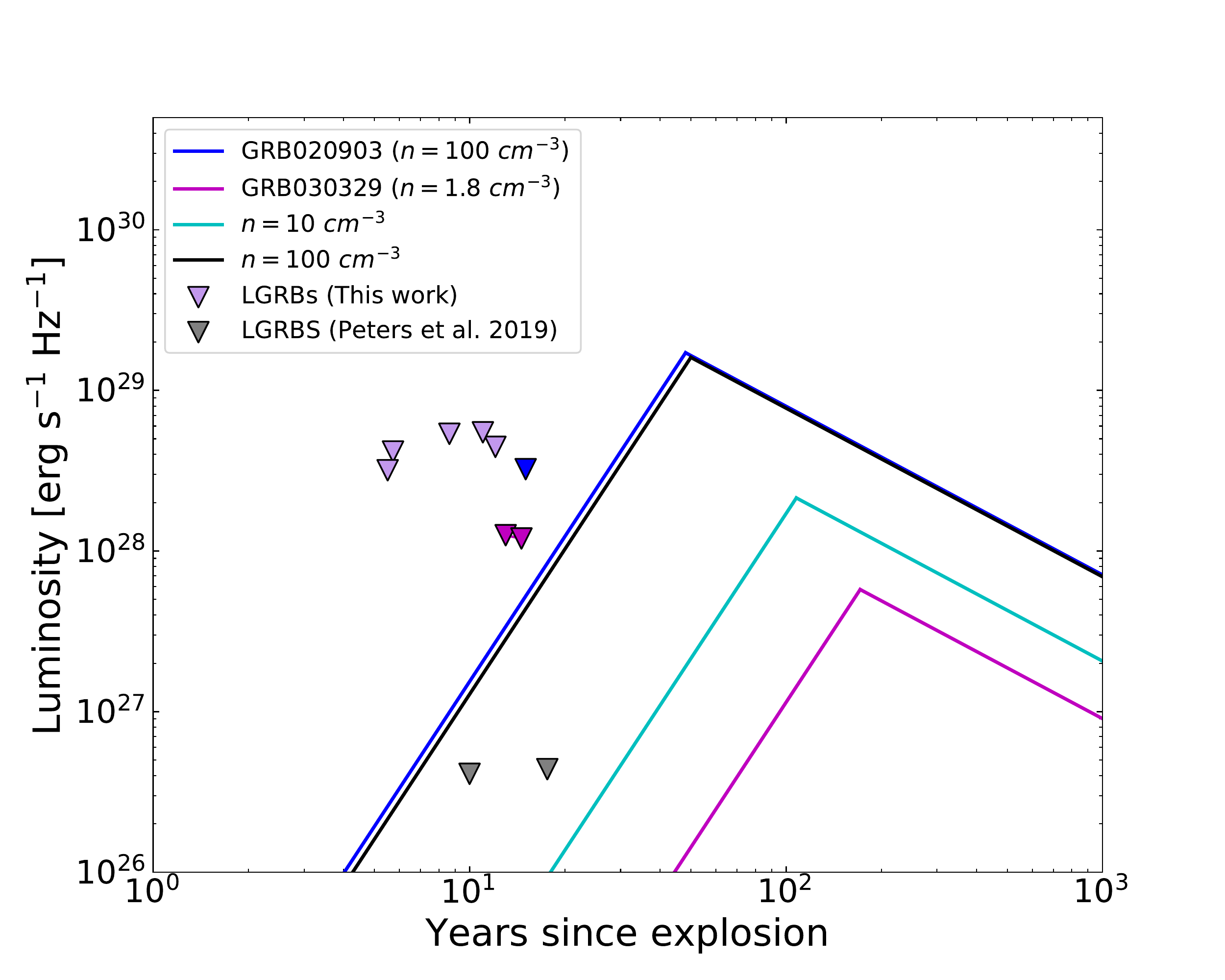}
\caption{Model lightcurves at 6 GHz for associated SNe in LGRBs assuming values of $\epsilon_e = \epsilon_B = 0.1$, $p=2.5$, $E_{\rm SN} = 10^{52}$ erg,  $v_{\rm ej} = 8 \times 10^3 \ \rm km \ s^{-1}$ and external number densities $n = 10$ and $n=100 \ \rm cm^{-3}$ compared to late-time limits presented in this work (light purple) and in \citet{Peters2019} (gray). For GRB\,020903 and GRB\,030329, we plot unique models with $n = 100$ (blue) and $n=1.8 \ \rm cm^{-3}$ (magenta) based on modeling of the GRB afterglows \citep{Berger2003b,Soderberg2004}.}
\label{fig:grb_sn}
\end{figure}

\begin{figure*}
\includegraphics[width=\textwidth]{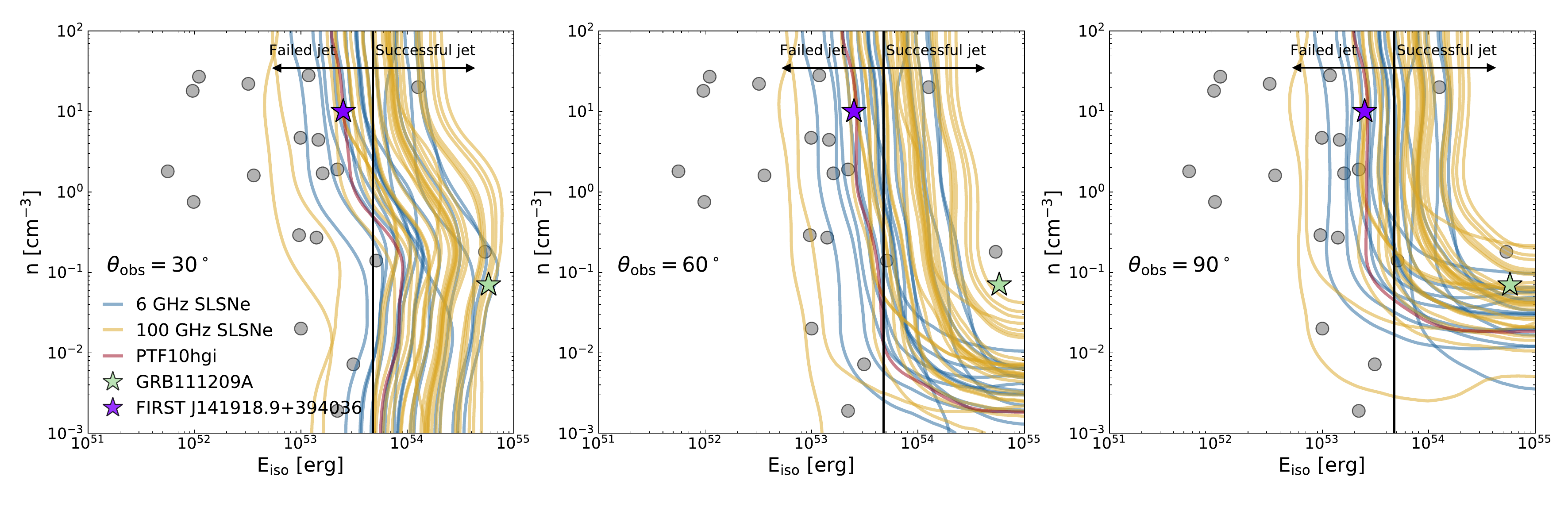}
\caption{Constraints on the allowed jet energies and CSM densities for off-axis jets with an initial opening angle $\theta_j = 10^\circ$ and viewing angles of $\theta_{\rm obs} = 30^\circ$, $\theta_{\rm obs} = 60^\circ$, and $\theta_{\rm obs} = 90^\circ$. Individual curves correspond to the allowed region of parameter space for each SLSN in our sample, where in each case the region to the right of the curve is ruled out by the 6 (blue) or 100 GHz (yellow) non-detection. Also shown are the range of jet energies and CSM densities consistent with the radio detection of PTF10hgi \citep{Eftekhari2019}. For comparison, we also include the extragalactic radio transient FIRST J141918.9+394036 \citep{Law2018} and the ultra-long GRB\,111209A \citep{Stratta2013}, as well as a sample of LGRBs (grey circles) from the literature \citep{Berger2001,Panaitescu2002,Berger2003b,Yost2003,Chevalier2004,Chandra2008,Cenko2010,Laskar2015}. The vertical black line depicts the median jet breakout energy $E_{\rm min}$ from Table~\ref{tab:magnetars}.}
\label{fig:jets}
\end{figure*}

\section{Radio Emission from External Shocks}
\label{sec:blastwave}

Radio observations of SNe provide a unique probe of the shock interaction between the ejecta and the surrounding CSM \citep{Chevalier1982}. This emission may arise from an initially off-axis relativistic jet that decelerates and spreads into the observer's line of sight at late times \citep{Rhoads1997,Sari1999}, or from the fastest layers of the SN ejecta \citep{Chevalier1998}. 

In the case of Type Ib/c SNe, for example, the radio emission can be used to track the mass-loss history of the progenitor star \citep{Soderberg2005,Soderberg2012,Wellons2012,Corsi2016,Palliyaguru2019}. The precise mechanism by which material is removed from the progenitor remains an open question, though is likely related to binary mass transfer, or for more massive stars, strong stellar winds \citep{Woosley1995}. In a small subset of events, there has been evidence for relativistic ejecta \citep{Soderberg2010,Margutti2014}.

The production of relativistic jets in LGRBs has been attributed to fallback accretion onto a black hole \citep{MacFadyen1999,MacFadyen2001} or the rotational energy of a millisecond magnetar (e.g., \citealt{Thompson2004,Metzger2011}). Recently, \citet{Margalit2018a} argued that similar jets may accompany SLSNe, if they are powered by magnetars. In this scenario, a misalignment of the magnetar's rotation and magnetic axes leads to a partitioning of the magnetar's spin-down luminosity, where some fraction of the spin-down power is converted into thermal energy behind the SN ejecta, while the remaining power energizes a relativistic jet. Thus, sufficiently energetic jets may break through the SN ejecta, producing late-time radio afterglow emission from SLSNe.

Below we use our radio limits to place constraints on the radio emission from external shocks due to both non-relativistic and relativistic outflows.

\subsection{Supernovae Ejecta}\label{sec:snejecta}

In the context of a shock interaction from the non-relativistic, quasi-spherical SN ejecta, we place limits on the shock velocity and mass-loss rate for each SLSN following the prescription of \citet{Chevalier1998} and assuming that the date of each observation corresponds to the peak luminosity; see Figure~\ref{fig:shock}. We also consider the possibility that the emission from each event may have peaked at earlier times, by extrapolating the peak luminosity back in time assuming $L_{\nu,\rm pk} \propto t^{-1}$ \citep{Berger2002}, and assuming that the emission is optically thin at the time of observations. This allows us to rule out the region of parameter space above each curve. 

We find that in this context, our limits at late times do not probe the range of mass-loss rates and ejecta velocities inferred for Type Ib/c SNe, which have mostly been detected at early time, with with typical inferred values of $v_{\rm ej}\sim 0.1$c and $A\sim 1-100 \ A^*$ (e.g., \citealt{Berger2002,Soderberg2005,Soderberg2012}). On the other hand, a small number of SLSNe limits from the literature at earlier times probe the typical ejecta velocities and mass-loss rates inferred for Type Ib/c SNe, and thus suggest that comparable outflows are not ubiquitous in SLSNe. One possible exception is PTF10hgi, in which the radio detection would imply an ejecta velocity and progenitor mass loss rate a few times larger than for Type Ib/c SNe. However, our interpretation for this event is that the radio emission is due to central engine activity.

For the LGRBs in our sample, we place constraints on emission from associated supernovae following the prescriptions of \citet{Barniol2015} and \citet{Kath2016}. In this scenario, radio emission is produced by a spherical outflow from the accompanying supernova which peaks at the deceleration time ($t_{\rm dec}$), when the SN has swept up a mass comparable to the initial ejected mass. During this phase, the light curve rises as\footnote{This is for the idealized scenario where the ejecta expands with a single velocity $v_{\rm ej}$ (see \citealt{Kath2016} for an extension to this scenario).}
$\propto t^3$ and decays after the deceleration time as $\propto t^{-3(1+p)/10}$. 
We assume typical values of $\epsilon_e = \epsilon_B = 0.1$ and $p=2.5$. We also include in our analysis late-time observations of three LGRBs from \citet{Peters2019}.

In Figure~\ref{fig:grb_sn}, we plot representative light curves for external number densities of $n=10$ and $n=100 \ \rm cm^{-3}$ and a fiducial supernova energy of $10^{52}$ erg, typical of broad-line Ic SNe associated with GRBs \citep{Mazzali2014}. For GRB\,020903 and GRB\,030329, we plot individual curves corresponding to the inferred densities ($n=100$ and $n=1.8 \ \rm cm^{-3}$, respectively) from modeling of the GRB afterglows \citep{Berger2003b,Soderberg2004}. We further fix the ejecta velocity to $8 \times 10^{3} \ \rm km \ s^{-1}$, consistent with the median observed nebular phase velocity for several broad-line Ic SNe (e.g., \citealt{Mazzali2001,Mazzali2007_sn2002ap,Mazzali2007_sn2006aj}). We note that this is in contrast to \citet{Peters2019}, in which the authors use the SN ejecta velocity at $\sim 10$ days after explosion, corresponding to velocities $3 - 4$ times larger than the observed nebular velocities. This has a dramatic effect on the resulting light curves which scale as $(v_{ej}/c)^{(11+p)/2}$ at $t < t_{dec}$. Indeed, we find that the light curve luminosities are roughly an order of magnitude fainter than our LGRB limits on the same timescale assuming $n = 100 \ \rm cm^{-3}$, with the exception of deep limits for GRB\,060218 and GRB\,980425 from \citet{Peters2019} which rule out these models, but cannot rule out densities $\lesssim 10 \ \rm cm^{-3}$. Furthermore, imposing the lower inferred nebular phase velocities shifts the peak of the lightcurves as $t_{\rm dec} \propto (v_{ej}/c)^{-5/3}$ to $t_{\rm dec} \gtrsim 50$ yr, well beyond the timescale of our observations at $\delta t \sim 10$ yr.
Finally, we note that even for a more promising choice of SN parameters (large ejecta velocity and energy) our radio non-detections may still be unconstraining due to inhibition of the early ($\lesssim$decades) SN light-curve that is predicted to be caused by interaction of the GRB jet with the ISM \citep{Margalit&Piran20}.

We thus conclude that we cannot place strong constraints on associated SNe in our sample of LGRBs, but note that such sources may be detectable on timescales of a few decades if they occur in high density ($n \approx 100 \ \rm cm^{-3}$) environments. Lower densities will require deep radio observations on century timescales.




\begin{figure*}
\includegraphics[width=\textwidth]{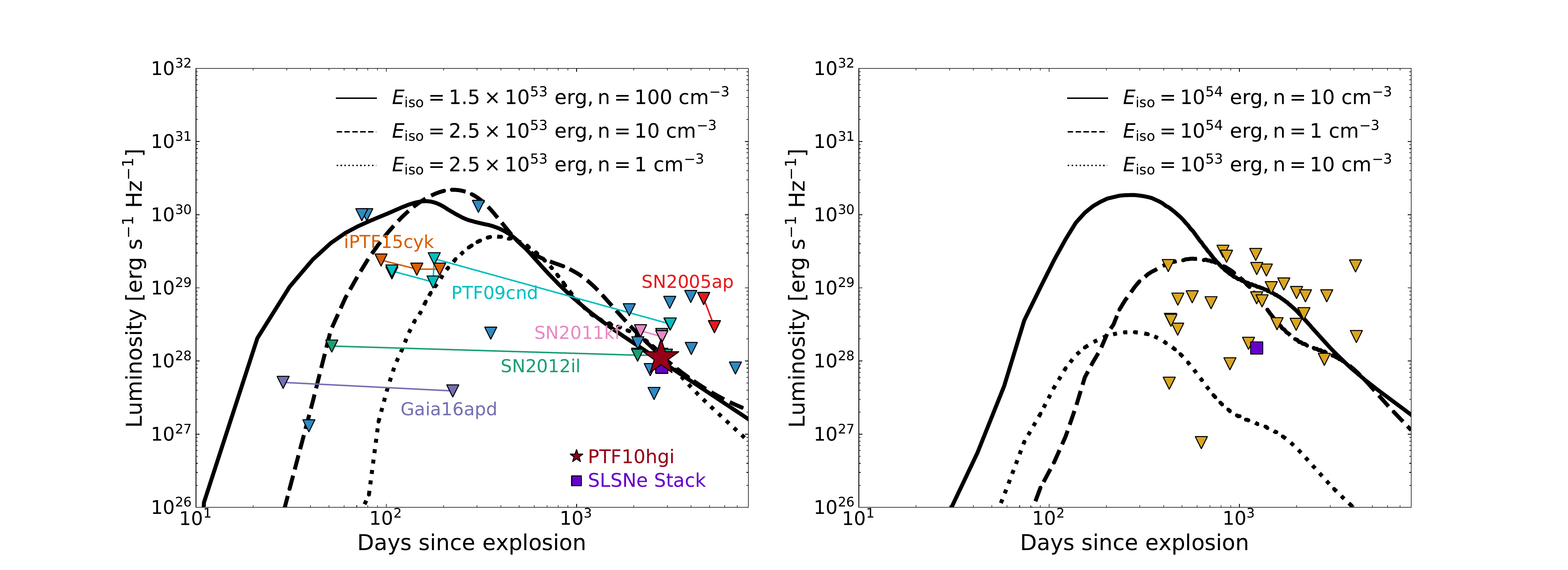}
\caption{\textit{Left}: Off-axis jet light curves at 6 GHz for $\theta_{\rm obs} = 60^\circ$ that are consistent with the radio detection of PTF10hgi. Individual curves depict a range of jet energies and CSM densities that can reproduce the observed flux density  at $\delta t \sim$ 7.5 years \citep{Eftekhari2019}. Shown for comparison are existing limits for SLSNe between 5 and 9 GHz (see Table~\ref{tab:litslsne}), where we highlight sources with multiple epochs of observations. \textit{Right}: Off-axis jet light curves at 100 GHz ($\theta_{\rm obs} = 60^\circ$) compared to upper limits for our sample of events. We also plot in both panels the luminosity corresponding to the stacked $3\sigma$ limit assuming the median observer time and redshift of the VLA and ALMA samples respectively.}
\label{fig:jetlcs}
\end{figure*}

\subsection{Off-axis Relativistic Jets}\label{sec:jets}

To constrain the presence of off-axis jets, we produce a grid of afterglow models for a range of jet energies and CSM densities using the two-dimensional relativistic hydrodynamical code {\tt Boxfit v2} \citep{vanEerten2012}. We generate afterglow light curves at 6 and 100 GHz and compare these to our upper limits to determine the allowed region of parameter space for each event. We model the light curves assuming a jet opening angle of $\theta_j = 10^\circ$ and viewing angles of $\theta_{\rm obs} = 30^\circ$, $60^\circ$, and $90^\circ$. We further assume a constant density CSM and jet microphysical parameters of $\epsilon_e = 0.1$, $\epsilon_B = 0.01$, and $p=2.5$, following previous studies of LGRBs (e.g., \citealt{Curran2010,Laskar2013,Wang2015,Laskar2016,Alexander2017}). 

The results are summarized in Figure~\ref{fig:jets}, where individual curves trace the jet energies and CSM densities corresponding to the $3\sigma$ flux density limit for each SLSN. In each case, the region of parameter space to the right of the curve (higher energies) is ruled out by the 6 and 100 GHz non-detections. We find that the radio limits largely preclude the presence of jets with $E_{\rm iso}\gtrsim 10^{54}$ erg ($n \sim 10^{-3} - 10^2 \ \rm cm^{-3}$) for $\theta_{\rm obs} = 30^\circ$ and $60^\circ$. For $\theta_{\rm obs} = 90^\circ$, we cannot rule out jets with $E_{\rm iso}\lesssim 10^{54}$ erg and $n \sim 10^{-3} - 10^{-2} \ \rm cm^{-3}$.

In Figure~\ref{fig:jetlcs}, we compare three representative off-axis jet light curves at 6 GHz to the limits for SLSNe, where we include all observations of SLSNe at $5-9$ GHz from the literature (see Table~\ref{tab:litslsne}). Individual curves depict jet energies of $E_{\rm iso} = 1.5 - 2.5 \times 10^{53}$ erg and CSM densities $n = 1 - 100 \ \rm cm^{-3}$, corresponding to a range of afterglow models that are consistent with the 6 GHz detection of PTF10hgi at $\delta t \sim 7.5$ yr \citep{Eftekhari2019}. We find that previous observations of SLSNe on timescales of a few hundred days post-explosion rule out the presence of jets with an energy scale similar to that inferred for PTF10hgi. Conversely, for the sample of SLSNe presented here we generally cannot rule out the presence of similar jets, primarily because most of the SLSNe are located at larger distances than PTF10hgi.

We also plot off-axis jet light curves at 100 GHz and compare to our ALMA upper limits in the right-hand panel of Figure~\ref{fig:jetlcs}. Because the peak of the afterglow emission is generically at GHz frequencies on such long timescales, our mm limits can accommodate higher jet energies of $E_{\rm iso}\gtrsim 10^{54}$ erg. This can also be seen in Figure~\ref{fig:jets}, where the ALMA limits trace out generally higher energies than the 6 GHz limits. We note that the two most constraining limits in the entire sample correspond to the lowest redshift sources in our sample, SN2018bsz and SN2018hti, which rule out off-axis jets with $E_{\rm iso}\gtrsim 10^{53}$ erg. 

We further rule out a population of off-axis jets with an energy scale significantly larger than is seen in typical LGRBs (Figure~\ref{fig:jets}). In this context, observations at earlier times ($\delta t \lesssim 1$ yr) are more constraining, as these probe the peak of the afterglow emission. 

As noted in \citet{Margalit2018a}, a successful jet will break out of the SN ejecta only if the jet head velocity exceeds the ejecta velocity, and if the kink stability criterion is satisfied, i.e., that magnetic energy is not dissipated due to kink instabilities \citep{Bromberg2016}. These conditions lead to a minimum threshold energy for successful jet breakout given by:
\begin{equation}
E_{\rm min} \simeq 0.195\, E_{K}  \bigg(\dfrac{\gamma_j}{2}\bigg)^{-4} f_j^{-1},
\end{equation}
where $E_K$ is the initial kinetic energy of the SN, $\gamma_j = 1/5 \theta_j$ as per \citet{Mizuta2013}, and $f_j$ is the fraction of energy contributing to a collimated jet \citep{Margalit2018a}. We assume $f_j = 0.5$ and adopt the inferred $E_K$ values for each source based on fits to the optical light curves. For each SLSN in our sample,\footnote{For this calculation we use only the sources with estimates of $E_K$ as derived from fits to the optical light curves using \texttt{MOSFIT} (see Table~\ref{tab:magnetars}).} we therefore estimate the minimal energy for jet breakout to determine whether our non-detections can constrain the theoretical predictions. The results are listed in Table~\ref{tab:magnetars}.

Among our VLA sources, we find that four SLSNe (SN2010kd, SN2011ke, SN2011kf, and SN2011kg) have radio limits that preclude the presence of a relativistic jet. That is, for the afterglow model parameters specified above, the radio limits probe jet energies below $E_{\rm min}$. Thus, a successful jet produced by one of these sources would be readily detected in our observations. The remaining VLA sources have jet breakout energies that are not ruled out by our observations, and thus may harbor jets that are below the sensitivity of our observations, but with $E_{\rm iso} \gtrsim E_{\rm min}$.

Similarly, among our ALMA sources, we can rule out the presence of off-axis jets at an observer viewing angle $\theta_{\rm obs}\lesssim30^\circ$ for 13 sources (SSS120810, SN2013dg, LSQ14bdq, LSQ14mo, OGLE15sd, SN2015bn, SN2016ard, SN2016els, SN2017gci, SN2017jan, SN2018hti, SN2018ibb, and SN2018lfe) given that the jet breakout energy is ruled out by our $3\sigma$ radio limits. For SN2016ard, we further rule out all jets with $\theta_{\rm obs} \lesssim 60^\circ$, as well as jets with $\theta_{\rm obs} \lesssim 90^\circ$ and $n \gtrsim 10^{-2} \ \rm cm^{-3}$. Similarly, for LSQ14bdq, SN2016els, and SN2018ibb we can rule out $\theta_{\rm obs} \lesssim 60^\circ$ and $n \gtrsim 10^{-2} \ \rm cm^{-3}$ and $\theta_{\rm obs} \lesssim 90^\circ$ and $n \gtrsim 10^{-1} \ \rm cm^{-3}$. For SN2017gci and SN2018hti, we can rule out all jets with $\theta_{\rm obs} \lesssim 60^{\circ}$ and jets with $\theta_{\rm obs} \lesssim 90^{\circ}$ and $n \gtrsim 10^{-1} \ \rm cm^{-3}$. Thus, for SN2016ard, LSQ14bdq, SN2016els, SN2018ibb, SN2017gci, and SN2018hti, the allowed phase-space for a successful jet that is not ruled out by our observations is prohibitively small. 

Finally, we note that the range of jet breakout energies ($E_{\rm min,iso} \approx 7 \times 10^{52} - 6 \times 10^{54}$ erg) for the SLSNe in our sample suggests that we do not expect a population of off-axis jets with energies comparable to the lowest energy LGRBs, e.g., $E_{\rm iso} \lesssim 10^{53}$ erg, given that such jets will fail to break out of the SN ejecta.

\begin{figure}
\center
\includegraphics[width=\columnwidth]{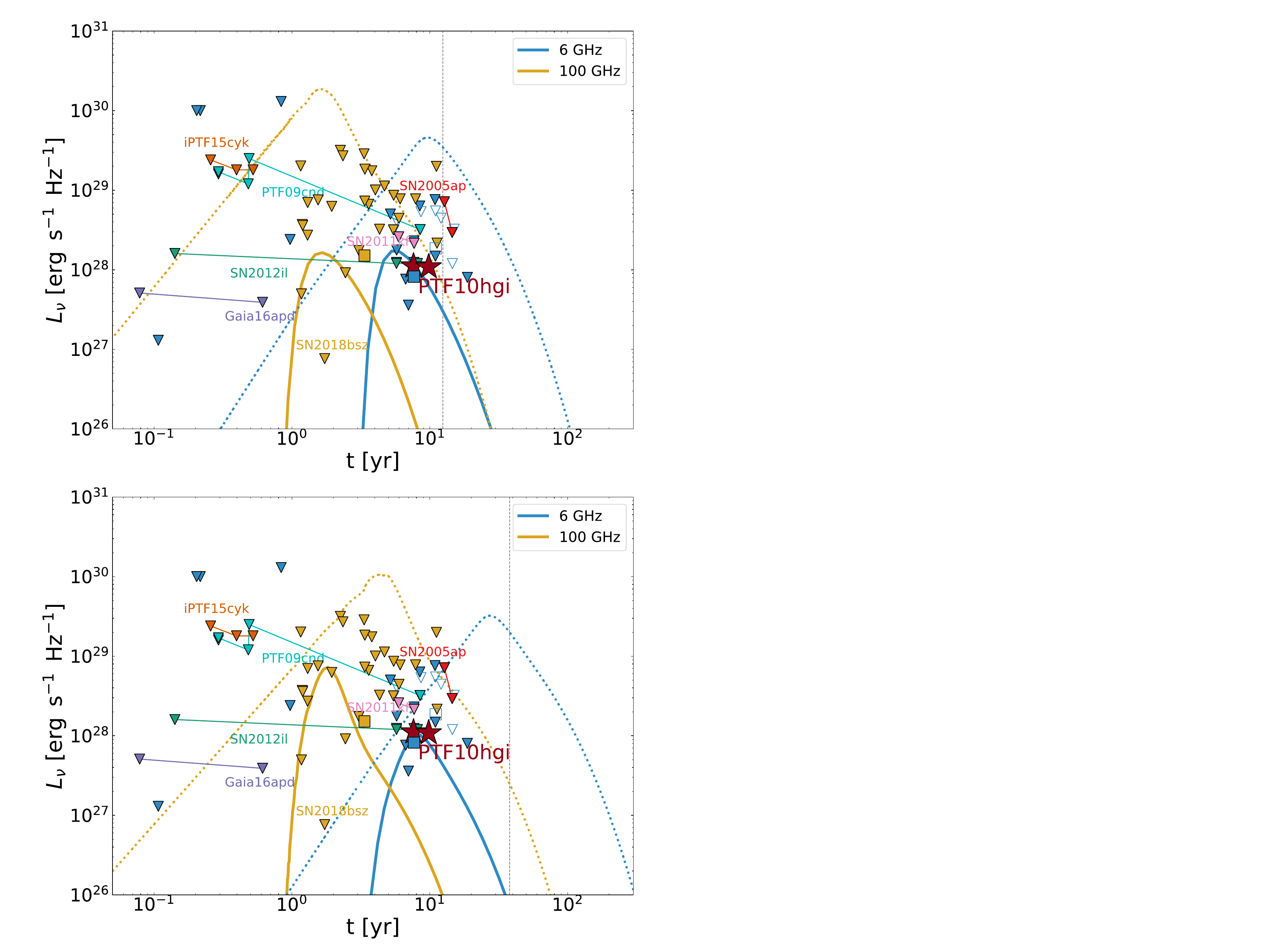}
\caption{Ion-electron nebula models at 6 GHz (blue) and 100 GHz (yellow) based on the prescription for \repeater{} from \citet{Margalit2018c} compared to upper limits on the radio luminosity for SLSNe, LGRBS, and the 6 GHz detection of PTF10hgi \citep{Eftekhari2019}. Yellow symbols correspond to our ALMA 100 GHz upper limits, blue filled symbols correspond to all limits for SLSNe between 5 and 9 GHz, and blue open symbols indicate LGRBs. The stacked $3\sigma$ luminosity for each population assuming the median observer time and redshift are plotted using square symbols. Also shown are existing limits for SLSNe between 5 and 9 GHz. The dashed curves correspond to models A (top panel) and B (bottom panel) used to describe \repeater{} from \citet{Margalit2018c}, where the vertical dashed line in each panel represents the presumed age of \repeater{} in each model. The solid lines depict models that are consistent with the PTF10hgi detection at 6 GHz. \textit{Top:} Model A from \citet{Margalit2018c} (dashed lines). Solid lines show the same model but with the magnetic energy scaled down by a factor of $\approx 20$ (i.e., $E_{\rm B_*} = 2.3 \times 
10^{49}$ erg) to explain the 6 GHz detection of PTF10hgi. \textit{Bottom:} Model B from \citet{Margalit2018c} (dashed lines). Solid lines correspond to a synchrotron self-
absorbed nebula that is consistent with the 6 GHz detection and 100 GHz limit for PTF10hgi as in \citet{Eftekhari2019}}.
\label{fig:nebula_models}
\end{figure}

\vspace{-0.4cm}
\section{Magnetar Wind Nebulae}\label{sec:nebula}

In the standard pulsar wind scenario, a newly formed neutron star acts as an energy reservoir for relativistic electron/positron pairs, leading to the production of a magnetized outflow, or pulsar wind nebula (PWN), inflated behind the expanding SN ejecta \citep{Gaensler2006}. Relativistic particles in the nebula are accelerated into a power-law distribution, producing non-thermal synchrotron radiation across the electromagnetic spectrum. In the GHz frequency regime, this emission is expected to penetrate the surrounding SN ejecta on timescales of several years to decades~\citep[][]{Murase2016,Omand2018}, as the ejecta becomes optically thin to free-free absorption and synchrotron self-absorption. 

\begin{figure*}
\vspace{-0.36cm}
\center\includegraphics[width=0.95\textwidth]{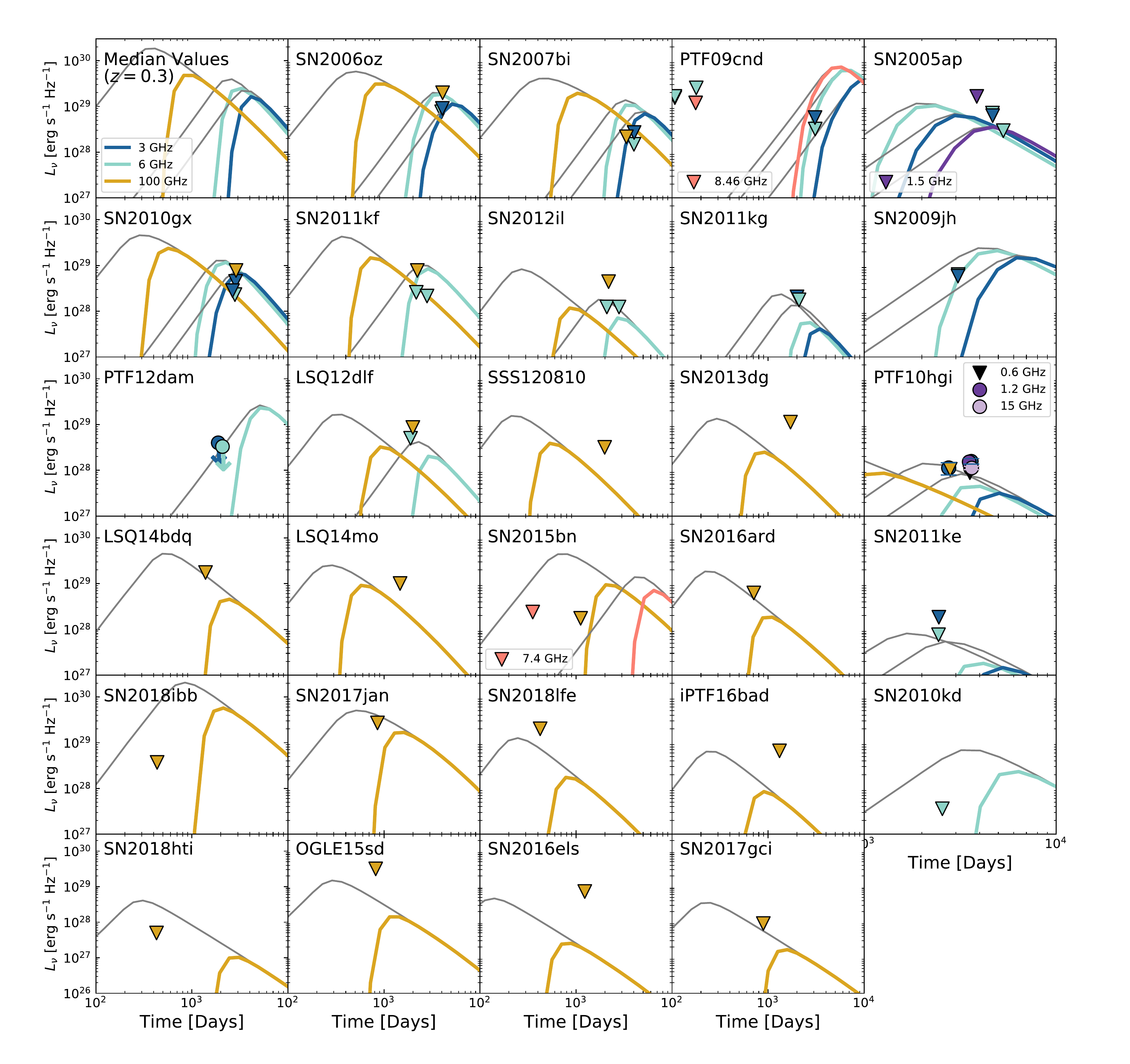}
\caption{Radio and millimeter light curves at 6 and 100 GHz for electron-positron PWN following the methods of \citet{Murase2016} and \citet{Omand2018}. Light curves are shown for all SLSNe with existing \texttt{MOSFiT} magnetar parameters as given in Table~\ref{tab:magnetars}. Gray curves indicate models with no absorption. The top left panel depicts the predicted radio light curve assuming the median magnetar parameters for the sample of SLSNe presented in \citet{Nicholl2017c}. In a few cases, we also include data at nearby frequencies from the literature (Table~\ref{tab:litslsne}), as well as models at 3 GHz for comparison to the models presented in \citet{Law2019}. For PTF12dam, we adopt the host galaxy detection values as an upper limit. Limits correspond to $3\sigma$.}
\label{fig:mag_lcs}
\end{figure*}

To model the PWN radio emission from each newborn SLSN remnant, we use the inferred magnetar birth properties (ejecta mass and velocity, magnetar magnetic field and spin period) from \citet{Nicholl2017b} and \citet{Blanchard2018}, and in a few cases presented here (see Table~\ref{tab:magnetars}). The parameters are derived via Markov chain Monte Carlo fits to the multicolor light curves using \texttt{MOSFiT}, an analytic Python-based modular code designed to model a variety of transients. We note that other approaches to modeling the optical light curves of SLSNe may lead to different parameter estimates (e.g., \citealt{Omand2018,Law2019}). Such studies use a spin-down formula based on numerical simulations \citep{Gruzinov2005,Spitkovsky2006,Tchekhovskoy2013} which gives a factor ${3(1+C\sin^2\chi_{\mu})/2\sin^2\chi_{\mu}} \sim 5$ larger spin-down luminosity for a given $P$ and $B_\perp$ ($\chi_{\mu}$ here is the angle between the magnetic and rotational axes and $C \sim 1$ is a numerical pre-factor) \citep{Kashiyama2016,Omand2018} . These models also allow for the pressure of the PWN to accelerate the SLSN ejecta, effectively coupling the ejecta kinetic energy to the spin-down luminosity of the magnetar. These approaches may lead to differences in the resulting light curves and parameters. We note however that a systematic comparison of these methods is beyond the scope of this work. 

Before computing the predicted PWN radio light curves for each source, we first assess the likelihood of detection for each SLSN in our sample\footnote{We perform this exercise exclusively for the sources listed in Table~\ref{tab:magnetars} as these events have the relevant engine and ejecta parameters from \texttt{MOSFIT} models.} by computing the free-free optical depth through the SN ejecta, and hence the transparency timescale for radio emission. Using the photoionization code \texttt{CLOUDY} \citep{Ferland2013}, we compute the time-dependent evolution of the temperature and ionization structure of the ejecta assuming that some fraction of the spin-down power of the magnetar is deposited into ionizing radiation. The ejecta density distribution for each SLSN is set by the inferred magnetar parameters. Assuming that the energy injection rate into the nebula is given by $L \propto t^{-2}$, we find $t_{\rm ff}\approx 1.3 - 16.7$ yr at 6 GHz and $\approx 1.2 - 4.6$ yr at 100 GHz. We therefore find that the majority of our sources are expected to be optically thin to radio emission at the time of our observations, with the exception of nine sources; SN2011kg and LSQ12dlf are expected to be optically thick at the time of our 6 GHz observations, while LSQ14bdq, SN2015bn, SN2016ard, SN2017jan, SN2018ibb, SN2018hti, and SN2018lfe are expected to be optically thick in our 100 GHz observations. This is consistent with the fact that these sources have larger inferred ejecta masses, and hence column densities, than the majority of SLSNe (with the exception of LSQ12dlf and SN2018lfe which fall below the median ejecta mass of the SLSNe distribution). Finally, we report upper limits on $t_{\rm ff, 100GHz}$ for eight sources where $t_{\rm ff, 100GHz} > 1 \rm{yr}$ $(1+z)$, as the \texttt{CLOUDY} models start from $t > 1$ yr in the source frame and these results are thus based on extrapolations. 

Next, we constrain the presence of nascent magnetar nebulae in our sources following two unique prescriptions. First, we consider the scenario presented in \citet{Margalit2018c} (see also \citealt{Beloborodov2017}), which posits an ion-electron wind produced by a young magnetar as the source of the radio emission for the quiescent counterpart associated with \repeater{} \citep{Kashiyama2017,Metzger2017}. Second, we consider an electron-positron wind, typical of standard PWNe, following the prescription of \citet{Omand2018}. In both cases, we use the engine and ejecta properties inferred from \texttt{MOSFiT} as described above.

\subsection{Ion-Electron Wind}
\label{sec:ion-wind}

\citet{Margalit2018c} proposed a magnetized ion-electron wind from a young magnetar to explain the observed properties of \repeater, namely the size and flux of the persistent radio counterpart, as well as the large and decreasing rotation measure (RM) of the bursts. The model assumes a one-zone nebula in which the magnetar's magnetic energy leads to the injection of relativistic electrons and ions that are thermalized at the termination shock of the magnetar wind. The model is characterized by the magnetic energy of the magnetar ($E_{B_*}$), the nebula velocity ($v_n$), the rate of energy input into the nebula ($t^{-\alpha}$) following the onset of the active period ($t_0$), the magnetization of the outflow ($\sigma$), and the mean energy per particle ($\chi$) assuming a proton-electron composition. An ion-electron plasma composition is invoked given that the RM contribution of an electron-positron plasma is zero and thus inconsistent with the large observed RM \citep{Michilli2018}. 

In Figure~\ref{fig:nebula_models}, we plot models A  and B from \citet{Margalit2018c}, corresponding to inferred source ages of $t_{\rm age}\approx 12.4$ and $38$ yr, respectively. In each case, the emission is expected to peak in the millimeter on timescales of a few years, and in the GHz regime on timescales of $\gtrsim 10$ yr. Moreover, the models predict 6 GHz luminosities of $\sim 10^{29} \ \rm erg \ s^{-1} \ Hz^{-1}$, consistent with the luminosity of the \repeater{} persistent radio source \citep{Chatterjee2017}. We find that our limits at both 6 and 100 GHz are sufficient to rule out these models, {\it with the same parameters as for \repeater}.

We further compare our limits to the models used to describe the 6 GHz radio detection and 100 GHz upper limit of PTF10hgi from \citet{Eftekhari2019} (see Figure~\ref{fig:nebula_models}). The first model is identical to model A for \repeater{} from \citet{Margalit2018c} with the magnetic energy scaled down by a factor of $\approx 20$, while the second model explores a scenario in which the 6 GHz emission is marginally self-absorbed. In \citet{Eftekhari2019}, we showed that a fully self-absorbed nebula (as constrained by the 100 GHz non-detection) is also consistent with the data. However, recent results presented by \citet{Mondal2020}
constrain the self-absorption frequency to $\sim 1$ GHz based on a steep drop in flux density between 0.6 and 1 GHz. Indeed, we find that the predicted light curve evolution of a marginally self-absorbed nebula is more consistent with the observed evolution of the 6 GHz luminosity.

We find that the majority of our limits are consistent with the model light curves invoked for PTF10hgi, although several deep limits at 100 GHz, most notably SN2018bsz with $L_\nu = 7.6\times 10^{26} \rm \ erg \ s^{-1} \ Hz^{-1}$, strongly rule out the presence of such nebulae. We note that SN2018bsz exhibits unusual properties, including a pre-maximum plateau in the optical light curve in addition to strong carbon lines, which suggest it may be an atypical event among SLSNe \citep{Anderson2018}. Although the timescale of our observations probe the predicted peak of emission, the sensitivity of our VLA and ALMA observations are not sufficient to rule out emission at the level of $L_\nu \lesssim 10^{28} \rm \ erg \ s^{-1} \ Hz^{-1}$. Thus, if ion-electron nebulae similar to the one inferred for PTF10hgi are ubiquitous in SLSNe, such sources will require deep radio observations, and may be more readily detected with future instruments such as the Square Kilometer Array (SKA).

It is worth noting however that PTF10hgi is among the nearest sources in our sample with $z = 0.0987$. Thus our non-detections for the remainder of events may be indicative of their distance. Only two sources are located at lower redshifts: SN2018hti and SN2018bsz with $z = 0.063$ and $z= 0.027$, respectively, both of which are observed at 100 GHz. In the case of SN2018hti, we find that the SN ejecta is optically thick to 100 GHz emission at the time of our observations, while magnetar parameters have not been derived for SN2018bsz. In this context, a non-detection for SN2018hti is thus unsurprising, while non-detections for the remaining events may simply point to their greater distance relative to PTF10hgi. 

Finally, the fact that the SLSNe radio limits preclude \repeater{} models suggests that there may be physical differences (e.g., magnetic field strength) between the magnetar engines responsible for powering SLSNe and that inferred for \repeater.

\begin{figure}
\includegraphics[width=\columnwidth]{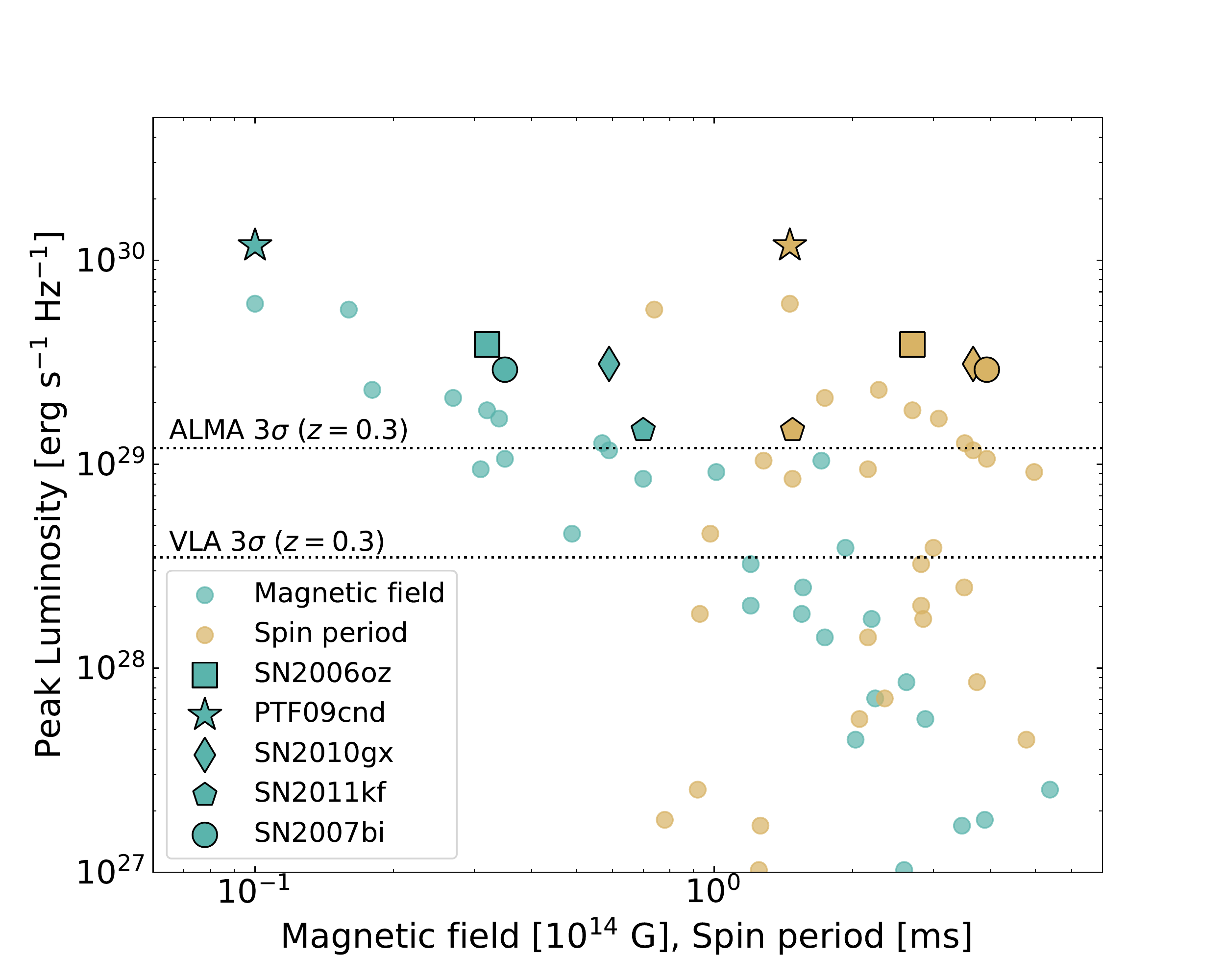}
\caption{Peak model luminosities (assuming maximal absorption) at 6 and 100 GHz based on the electron-positron PWN light curves in Figure~\ref{fig:mag_lcs} as a function of magnetar magnetic field and spin period.  We highlight individual sources that indicate predicted emission at or above the level of our $3\sigma$ limits at the time of our observations. Horizontal lines correspond to the typical $3\sigma$ sensitivity of our VLA and ALMA observations assuming $z = 0.3$.}
\label{fig:magnetar_corrs}
\end{figure}

\subsection{Electron-Positron Wind}
\label{sec:e-p-wind}

Here we consider a modified version of the PWN model presented in \citet{Omand2018} \citep[see also][]{Murase2020}, following the prescription of \cite{Murase2016} for quasi-steady radio emission from magnetar engines in SLSNe and FRBs. An electron-positron wind nebulae has also been considered to explain the observed properties of FRB 121102~\citep{Murase2016,Kashiyama2017}. In this work, we modify the spin-down formula and neutron star masses, and ignore the effects of ejecta feedback for consistency with the \texttt{MOSFiT} models. The radio models solve the Boltzmann equation for photons and electron/positrons in the PWN over all photon frequencies and electron energies \citep{Murase2015}, allowing for a self-consistent calculation of pair cascades, Compton and inverse Compton scattering, adiabatic cooling and both internal and external attenuation.  The electron-positron injection spectrum is assumed to be a broken power law with injection spectral indices of $q_1=1.5$ and $q_2=2.5$ and a peak Lorentz factor of $\gamma_{b} = 10^{5}$, which is consistent with Galactic PWNe (e.g., \citealt{Tanaka2010,Tanaka2013}) such as the Crab PWN, as well as the inferred nebula for PTF10hgi with $q_1 = 1.3$ \citep{Mondal2020}. Free-free absorption in the ejecta is calculated assuming a singly-ionized oxygen ejecta, and we do not consider absorption outside the ejecta, as in \citet{Omand2018} and \citet{Law2019}. 

The results of the models are shown in Figure~\ref{fig:mag_lcs} where we plot the predicted light curves at 6 and 100 GHz for the SLSNe in our sample with existing \texttt{MOSFiT} parameters (Table~\ref{tab:magnetars}). The models generically predict emission which peaks initially in the millimeter with an increased flux density on timescales of $\sim 1000$ days and cascades to lower frequencies and lower flux densities at later times. In this context, the timescales of our ALMA observations lead to less constraining limits as they do not probe the peak of the emission, which is expected at earlier times. In a few cases, we also include models at 3 GHz for comparison to limits and models presented in \citet{Law2019}. We generate new 3 GHz models using the formalism described above for self-consistency.

For the majority of our sources, we find that the non-detections are consistent with the predicted light curves. At 6 GHz, a number of sources are expected to peak at later times relative to the timescale of our observations, and hence may be detected in the future with continued monitoring. We note that five sources exhibit predicted emission at or above the level of our $3\sigma$ limits at the time of our observations. This includes SN2006oz, PTF09cnd and SN2011kf at 6 GHz, SN2010gx at 3 and 6 GHz, and SN2007bi at 3, 6, and 100 GHz. For an additional six sources (SN2009jh, SN2010kd, SN2012il, SN2017jan, SN2018ibb, and SN2018hti), the limits exclude models without absorption but cannot rule out models with absorption. 

In Figure~\ref{fig:magnetar_corrs}, we plot the predicted peak luminosity for each SLSNe (assuming maximal absorption) as a function of the inferred SLSN magnetic field and spin period. We find a strong anti-correlation between the peak luminosity and magnetic field, i.e., the more luminous events correspond to lower magnetic fields. This is consistent with the fact that the assumed pulsar spin-down luminosity $L_{\rm em} \propto B^{-2}$ as per numerical simulations \citep{Gruzinov2005}. Indeed, the five sources with predicted emission at or above the level of our $3\sigma$ limits correspond to exclusively low-$B$  ($B\lesssim 10^{14}$ G) events. Interestingly, from Figure~\ref{fig:magnetar_corrs}, we also find that our VLA and ALMA observations are sensitive to a large fraction of low-$B$ events, and thus indicate that continued monitoring on the appropriate timescales may reveal emission from these sources. Conversely, we find no clear correlation between the peak luminosity and the spin period. 

\begin{figure}
\includegraphics[width=\columnwidth]{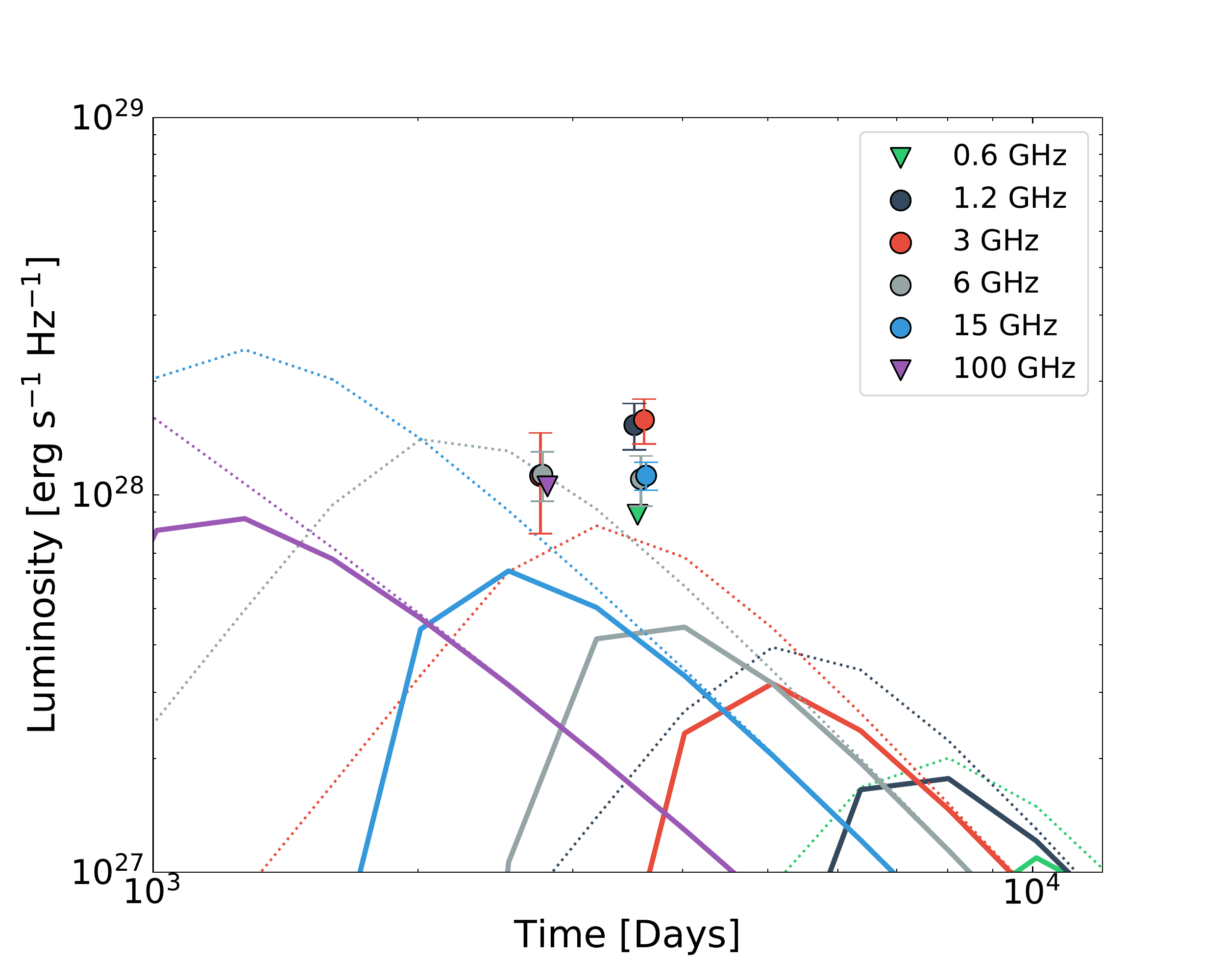}
\caption{Electron-positron nebula model light curves for PTF10hgi at $0.6-100$ GHz (\S\ref{sec:e-p-wind})  compared to the data from \citet{Eftekhari2019,Law2019,Mondal2020}. Dashed curves correspond to models without absorption.}
\label{fig:10hgi}
\end{figure}

For PTF10hgi, we plot the light curves over the range $0.6 - 100$ GHz in Figure~\ref{fig:10hgi}. We find that the observations at 6 GHz at $\delta t \approx 8$ yr \citep{Eftekhari2019} are consistent with the model with no absorption, while the same models underpredict the data at $\delta t \approx 10$ yr \citep{Mondal2020}. At 3 GHz, the models slightly underpredict the data at $\delta t \approx 8$ yr \citep{Law2019}, however, this difference is negligible in light of systematic uncertainties. The 3 GHz light curve is predicted to rise between $\delta t \approx 8 - 10$ yr, consistent with the shape of the observed data, but with a scale significantly below the observations. Non-detections at both 0.6 and 100 GHz are consistent with the models. Conversely, the data at 1.2 and 15 GHz lie well above the predicted light curves.

In \citet{Law2019}, the authors use unique magnetar parameters for PTF10hgi ($B = 1.4 \times 10^{14}$ G, $P = 1$ ms, $M_{\rm ej} = 15 \ M_{\odot}$ based on by-eye estimates), and find that a model in which $\sim 40\%$ of the ejecta is singly ionized is consistent with the observed data. However, such a model fails to accurately predict the latest SED obtained in \citet{Mondal2020}. Namely, the models slightly underpredict the data at $1.2 - 6$ GHz, but vastly underpredict the data at 15 GHz. This is in contrast to the models presented here in which the low frequency model at 1.2 GHz drastically underpredicts the observed flux value. This suggests that the true magnetar parameters may lie somewhere between the inferred values in both models (e.g., $P \approx 2 -4$ ms, $M_{\rm ej} \approx 6 - 9 \ M_{\odot}$). On the other hand, the flat shape of the SED at $1-15$ GHz may indicate a steeper injection spectrum, or an ionization fraction that evolves as a function of time. Continued broadband radio monitoring will provide a test of these scenarios. 

\begin{figure}
\includegraphics[width=\columnwidth]{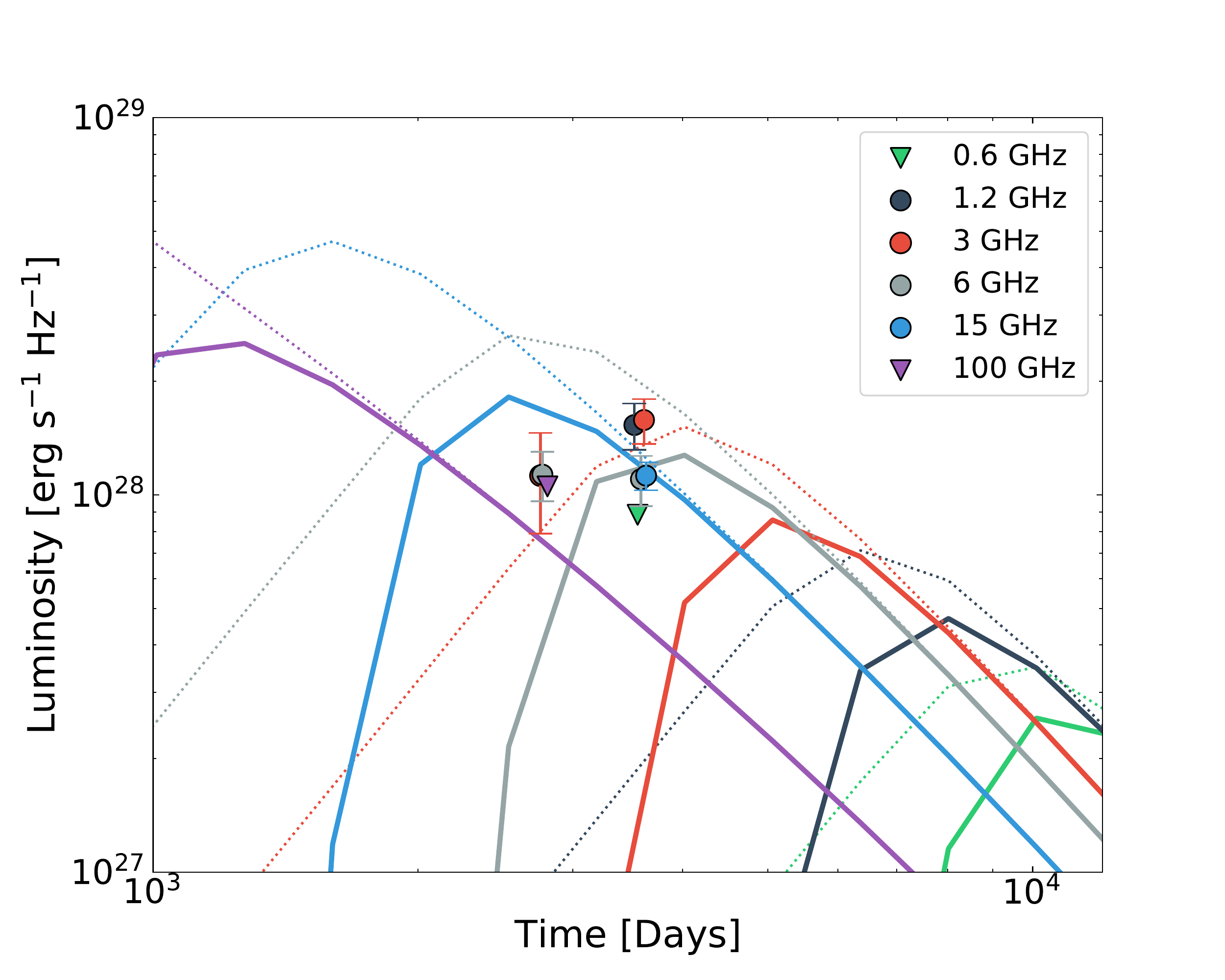}
\caption{Electron-positron nebula models for PTF10hgi as in Figure~\ref{fig:10hgi}, but with a peak Lorentz factor of $\gamma_b = 10^4$.}
\label{fig:10hgi_gb1e4}
\end{figure}

In Figure \ref{fig:10hgi_gb1e4} we plot the same data and models, but with a peak Lorentz factor of $\gamma_b = 10^4$ instead of $10^5$; we find these lightcurves are more consistent with the data.  Specifically, we find that the models are below the non-detection upper limits at 0.6 and 100 GHz, and are consistent with the 3 GHz observations assuming no absorption and with the 6 GHz observations assuming partial absorption at $\delta t \approx 6$ yr and almost maximal absorption at $\delta t \approx 8$ yr. The models are consistent with the data at 15 GHz regardless of absorption (the ejecta is predicted to be optically thin for free-free absorption at 15 GHz at the relevant timescale). Conversely, the model vastly underpredicts the 1.2 GHz data which is predicted to peak at later times.


One interesting physical implication of decreasing $\gamma_b$ is that the pair multiplicity in PTF10hgi is significantly higher than that of the Crab pulsar or other Galactic PWNe, which may indicate that the pair formation or acceleration processes in the nebulae of supernovae driven by highly magnetic millisecond pulsars are qualitatively different then those of Galactic PWNe. Neverthless, it is unclear whether this is unique to SLSNe nebulae due to the luminosity of the nebula and strength of the pulsar field, or if SLSNe nebulae eventually evolve to have a lower multiplicity.
  

\section{Conclusions}
\label{sec:conc}

We presented the largest and deepest sample of radio and millimeter observations of SLSNe to date. Using the VLA and ALMA, our observations probe non-thermal synchrotron emission from these SLSNe, as well as a small sample of nearby LGRBs, on timescales of $\sim 1 - 19$ years post explosion. Combined with existing observations from the literature, we place constraints on obscured star formation in the host galaxies, non-relativistic ejecta, central engines, and the possible connection to FRBs. Our key results are summarized as follows: 

\begin{itemize}
\item{We do not detect FRBs from any of our sources, placing limits on the maximum burst energies of $E_{\rm b,max}\lesssim 2 \times 10^{37} - 4 \times 10^{38}$ erg, comparable to the lowest energy bursts detected from \repeater{} \citep{Gourdji2019}. However, the likelihood of a detection in such short duration observations ($\sim 40$ min) is low, particularly in light of the intermittent nature of \repeater. Thus further follow-up and a larger time investment with the GBT, Arecibo, or FAST is warranted.}

\item{We find no evidence for significant dust obscuration within the host galaxies. PTF12dam is the only host that is detected in the radio, with a corresponding SFR of $\approx 7 \rm \ M_{\odot} \ yr^{-1}$, consistent with the results presented in \citet{Hatsukade2018}, and comparable to the H$\alpha$ inferred rate of $\approx 5 \ \rm M_{\odot} \ yr^{-1}$. Thus there is no indication of significant dust attenuation within the host galaxy. For the remainder of the SLSN hosts, we find that the radio-inferred SFR upper limits exceed optical estimates by factors of $\approx 2 - 5 \times 10^3$. Although some of our limits do not provide meaningful constraints in this context, for the majority of events, we can rule out significant dust obscuration in the SLSNe host galaxies. This is consistent with previous findings that indicate an absence of highly-reddened SLSNe \citep{Schulze2018}.}

\item{In the context of non-relativistic outflows (in analogy with Type Ib/c SNe), our late-time limits place direct constraints on low ejecta velocites ($v_{\rm ej} \approx 10^2 - 10^3\ \rm km \ s^{-1}$) and large CSM densities ($A \approx 10^3 - 10^5 A^*$). Extrapolating our limits back to earlier times (when Type Ib/c SNe are typically detected), we find that the inferred radio luminosities are roughly an order of magnitude larger than radio-detected Type Ib/c SNe. This suggests that SLSNe may indeed produce non-relativistic outflows typical of Type Ib/c SNe. Such constraints will require deep radio observations of SLSNe at earlier times, of which there are currently only a small number.}

\item{For the LGRBs in our sample, we constrain the emission from associated supernovae which is expected to peak at late times. We find that the low nebular phase velocities of broad-line Ic SNe imply faint radio emission at the level of a few $\times 10^{27} - 10^{29} \ \rm erg \ s^{-1} \ Hz^{-1}$ which peaks on timescales of several decades to centuries, well beyond the timescale of our observations. We therefore cannot place strong constraints on associated SNe in our sample of LGRBs, but note that such sources may be detectable at later times, particularly if they occur in high density ($n \approx 100 \ \rm cm^{-3}$) environments. }

\item{Our constraints on emission from off-axis relativistic jets allow us to rule out jets with $E_{\rm iso}\gtrsim 10^{54}$ erg for $\theta_{\rm obs}\lesssim 60^{\circ}$ and a wide range of CSM densities ($n\sim 10^{-3} - 10^{2} \ \rm cm^{-3}$). For extreme off-axis viewing angles ($\theta_{\rm obs} \sim 90^\circ$), we cannot rule out energetic jets with $E_{\rm iso}\gtrsim 10^{54}$ and low densities of $n\sim 10^{-3} - 10^{-2} \ \rm cm^{-3}$. We also estimate the minimum energy required for a jet to break out of the SN ejecta from each source and find that the range of breakout energies, $E_{\rm min,iso}\approx 7 \times 10^{52} - 6 \times 10^{54}$ erg, precludes a population of off-axis jets with energies comparable to the lowest energy LGRBs ($E_{\rm iso} \lesssim 10^{53}$ erg), as such jets will fail to break out of the SN ejecta.}

\item{Our limits are sufficient to rule out the presence of ion-electron magnetar nebulae with the same parameters invoked for \repeater, as they probe luminosities roughly an order of magnitude deeper. On the other hand, we find that the majority of our limits are consistent with the same models tuned to the radio detection of PTF10hgi. This may point to some differences (e.g., magnetic field strength) between the magnetar engine powering \repeater{} and those found in SLSNe.}

\item{In the context of an electron-positron PWN with an energy injection spectrum based on Galactic PWNe, we find that the majority of our non-detections are consistent with the predicted light curves which peak at later times or are below the nominal sensitivity of our observations. On the other hand, the models for a small number of events predict emission at or above the level of our limits. These events correspond to exclusively low-$B$ sources, with $B \lesssim 10^{14}$ G. Such discrepancies may reflect a time-varying ionization state or point to uncertainties in the inferred magnetar parameters.}

\item{A standard electron-positron PWN is unable to explain the observed spectrum for PTF10hgi. Namely, using the inferred engine parameters presented here, the light curve at 1.2 GHz significantly underpredicts the observed data. This is in contrast to the models presented in \citet{Law2019} which use unique engine parameters and subsequently underpredict the data at 15 GHz. This suggests that the true engine parameters may lie somewhere between the inferred values in both models. Conversely, the ionization fraction of the ejecta may evolve as a function of time, or the energy injection spectrum may be steeper than assumed. We further decrease the peak Lorentz factor to $\gamma_b = 10^4$ and find that the resulting light curves are somewhat more consistent with the data, yet still unable to explain the full SED of PTF10hgi. A decrease in $\gamma_b$ would imply a significantly higher pair multiplicity in PTF10hgi relative to Galactic PWNe.}
\end{itemize}

Continued radio monitoring of these sources on timescales of $5-10$ yr may lead to detections, particularly in the context of PWNe models which predict that the 6 GHz emission will peak at even later times. At 100 GHz, improved constraints will require both earlier observations and in many cases increased sensitivity, to probe the levels of emission associated with PWNe. In this context, targeting nearby SLSNe with small inferred B-values may provide the fastest route to a detection. Conversely, constraints on jetted emission predicate the need for multiple epochs of observations, as these are more constraining on the allowed region of parameter space with respect to observer angles, jet energies, and CSM densities. Finally, continued searches for FRBs from these sources may lead to detections, establishing a connection between FRBs and SLSNe and/or LGRBs. 

\software{Boxfit \citep{vanEerten2012}, CASA \citep{McMullin2007}, CLOUDY \citep{Ferland2013},  MOSFIT\citep{Guillochon2018},  pwkit \citep{Williams2017}}

\acknowledgments \textit{Acknowledgments.} The Berger Time-Domain Group at Harvard is supported in part by the NSF under grant AST-1714498. SC and JMC acknowledge support from the National Science Foundation (AAG-1815242). MN is supported by a Royal Astronomical Society Research Fellowship. CMBO has been supported by the Grant-in-aid for the Japan Society for the Promotion of Science (18J21778). KDA acknowledges support provided by NASA through the NASA Hubble Fellowship grant HST-HF2-51403.001 awarded by the Space Telescope Science Institute, which is operated by the Association of Universities for Research in Astronomy, Inc., for NASA, under contract NAS5-26555. RM acknowledges partial support by the National Science Foundation under Award No. AST-1909796 and AST-1944985 and by the Heising-Simons Foundation under grant \# 2018-0911. RM is a CIFAR Azrieli Global Scholar in the Gravity \& the Extreme Universe Program, 2019 and an Alfred P. Sloan Fellow in Physics, 2019. BM is supported by NASA through the NASA Hubble Fellowship grant \#HST-HF2-51412.001-A awarded by the Space Telescope Science Institute, which is operated by the Association of Universities for Research in Astronomy, Inc., for NASA, under contract NAS5-26555. The VLA observations presented here were obtained as part of programs VLA/17B-171, PI: Berger and VLA/19A-295, PI: Eftekhari. The VLA is operated by the National Radio Astronomy Observatory, a facility of the National Science Foundation operated under cooperative agreement by Associated Universities, Inc. This paper makes use of the following ALMA data: ADS/JAO.ALMA\#2017.1.00280.S., ADS/JAO.ALMA\#2019.1.01663.S. ALMA is a partnership of ESO (representing its member states), NSF (USA) and NINS (Japan), together with NRC (Canada), MOST and ASIAA (Taiwan), and KASI (Republic of Korea), in cooperation with the Republic of Chile. The Joint ALMA Observatory is operated by ESO, AUI/NRAO and NAOJ.

\vspace{-0.2cm}
\startlongtable
\begin{deluxetable*}{lcccccc}
\tablecolumns{7}
\tablecaption{Existing Radio Observations of SLSNe from the Literature.}
\tablehead{
\colhead{Source} & 
\colhead{$z$} & 
\colhead{Explosion Date} & 
\colhead{$\delta t^{\rm a}$} & 
\colhead{Frequency} & 
\colhead{Flux} &
\colhead{Reference} \\ 
\colhead{} & 
\colhead{} & 
\colhead{[MJD]} & 
\colhead{[days]} &
\colhead{[GHz]} & 
\colhead{[$\mu$Jy]} & 
\colhead{}
}  
\startdata
SN2005ap & 0.2832 & 53424& 3009 & 1.5 & ${<75}$ & \citet{Schulze2018} \\
...&... &... & 3613, 3668$^\ddagger$ & 3 & $<30$ & \citet{Law2019}\\
SN2006oz & 0.376 & 54025 & 2915, 2979$^\ddagger$ & 3 & $<24$ & \citet{Law2019} \\
SN2007bi & 0.1279 & 54075 & 3546, 3593$^\ddagger$ & 3 & $<66$ & \citet{Law2019} \\
SN2009jh & 0.35 & 55014 & 2256, 2309$^\ddagger$ & 3 & $<18$ & \citet{Law2019} \\
PTF09atu & 0.5015 & 55025$\rm ^{b}$ & 2011, 2068$^\ddagger$ & 3 & $<24$ & \citet{Law2019} \\
PTF09cnd & 0.258 & 55006 & 85 & 8.46 & $<90$ & \citet{Chandra2009} \\ 
...& ...& ...& 85 & 4.86 & $<96$ & \citet{Chandra2009} \\ 
...& ...& ...& 85 & 1.41 & $<579$ & \citet{Chandra2009} \\ 
...& ...& ...& 140 & 8.46 & $<69$ & \citet{Chandra2010} \\ 
...& ...& ...& 142 & 4.86 & $<147$ & \citet{Chandra2010} \\ 
...& ...& ...& 147 & 1.41 & $<600$ & \citet{Chandra2010} \\ 
...& ...& ...& 2416, 2483$^\ddagger$ & 3 & $<33$ & \citet{Law2019}\\
PS1-10awh & 0.908 & 55467 & 39 & 4.9 & $<45$ & \citet{Chomiuk2011} \\
PS1-10bzj & 0.65 & 55523$^{\rm c}$& 48 & 4.96 & $<87$ & \citet{Coppejans2018} \\
PS1-10ky & 0.956 & 55299$^{\rm d}$& 156 & 4.9 & $<51$ & \citet{Chomiuk2011} \\
PTF10aagc & 0.206 & 55439$^{\rm e}$ & 2041 & 3 & $<15.6$ & \citet{Hatsukade2018} \\
PTF10hgi & 0.0987 & 55322$^{\rm f}$ & 2524 & 6 & 47.3 & \citet{Eftekhari2019} \\
... & ... & ... & 2469, 2547$^\ddagger$ & 3 & $47$ & \citet{Law2019} \\
... & ... & ... & 3220, 3250$^\ddagger$ & 0.6 & $<36$ & \citet{Mondal2020}\\
... & ... & ... & 3207 & 1.2 & 64 & \citet{Mondal2020}\\
... & ... & ... & 3289 & 3.3 & 66 & \citet{Mondal2020}\\
... & ... & ... & 3263 & 6 & 46 & \citet{Mondal2020}\\
... & ... & ... & 3308& 15 & 47 & \citet{Mondal2020}\\
PTF10uhf & 0.288 & 55388$^{\rm g}$ & 1951 & 3 & $<80.9^\dagger$ & \citet{Hatsukade2018} \\
SN2010gx & 0.230 & 55251& 2154 & 3 & $<20.7$ & \citet{Hatsukade2018} \\
...& ...& ...& 2295, 2339$^\ddagger$ & 3 & $<33$ & \citet{Law2019}\\
SN2010kd & 0.1 & 55499 & 2341, 2390$^\ddagger$ & 3 & $<42$ & \citet{Law2019} \\
SN2011kg & 0.192 & 55912 & 1678 & 3 & $<22$ & \citet{Hatsukade2018} \\
SN2011ke & 0.14 & 55651 & 2113, 2175$^\ddagger$ & 3 & $<36$ & \citet{Law2019} \\
SN2012il & 0.175 & 55919 & 44 & 5.9 & $<21$ & \citet{Chomiuk2012} \\
PTF12dam & 0.107 & 56022 & 1697 & 3 & $<141.5^\dagger$ & \citet{Hatsukade2018} \\
iPTF15cyk & 0.539 & 57249$^{\rm h}$& 61 & 5.4 & $<30$ & \citet{Kasliwal2016} \\
...&... &... & 94 & 5.4 & $<23$ & \citet{Kasliwal2016} \\
...&... &... & 124 & 5.4 & $<23$ & \citet{Kasliwal2016} \\
SN2015bn & 0.1136 &57013 & 318 & 7.4 & $<75$ & \citet{Nicholl2016b} \\
...&... & ...& 318 & 22 & $<40$ & \citet{Nicholl2016b} \\
...&... & ...& 867 & 21.8 & $<48$ & \citet{Nicholl2018} \\
...&... & ...& 867 & 33.5 & $<63$ & \citet{Nicholl2018} \\
Gaia16apd & 0.1013 & 57512$^{\rm i}$& 26 & 5.9 & $<20.4$ & \citet{Coppejans2018} \\ 
... &... &... & 26 & 21.8 & $<45.9$ & \citet{Coppejans2018} \\ 
...&... &... & 203 & 5.9 & $<15.3$ & \citet{Coppejans2018} \\ 
...&... &... & 203 & 21.8 & $<30.1$ & \citet{Coppejans2018} \\ 
SN2017egm & 0.0307 & 57887$^{\rm j}$ & 34 & 15.5 & $<1800^\dagger$ & \citet{Bright2017} \\ 
...&... &... & 38 & 5 & $<60$ & \citet{Bright2017} \\ 
...&... &... & 39 & 10 & $<76.8$ & \citet{Bose2018} \\ 
...&... &... & 39 & 1.6 & $<96$ & \citet{Bose2018} \\ 
...&... &... & 46 & 33 & $<150$ & \citet{Coppejans2018} \\ 
...&... &... & 47 & 10 & $<86.1$ & \citet{Bose2018} \\ 
...&... &... & 74 & 33 & $<33.6$ & \citet{Coppejans2018} \\ 
\enddata
\tablecomments{Limits correspond to 3$\sigma$.\\
$^{\rm a}$ Explosion rest-frame\\
$^{\rm b}$ \citet{DeCia2018}.\\
$^{\rm c}$ Assuming a fast rise rest-frame time of 25 days and a peak time of MJD 55563.65$\pm 2$ \citep{Lunnan2018}. \\
$^{\rm d}$ Assuming a rest-frame rise time of 50 days and a peak time of MJD 55397 \citep{Chomiuk2011}.\\
$^{\rm e}$ Assuming a rest-frame rise time of 50 days and a peak time of MJD 55499.5 \citep{Quimby2018}.\\
$^{\rm f}$ \citet{Inserra2013}.\\
$^{\rm g}$ Assuming a rest-frame rise time of 50 days and a peak time of  MJD 55452.3 \citep{Quimby2018}.\\
$^{\rm h}$ We assume a peak time of MJD 57293.5 and a rest-frame rise time of 60 days as in \citet{Coppejans2018}.\\
$^{\rm i}$ From \citet{Yan2015}, the peak time is MJD 57541  and the rest-frame rise time is 29 days \citep{Nicholl2017d}.\\
$^{\rm j}$ \citet{Nicholl2017a}.\\
$^{\dagger}$ The host galaxy contribution to the flux density is unknown, so we adopt their detection value as an upper limit.\\
$^{\ddagger}$ $3\sigma$ limit averaged over two epochs.}
\label{tab:litslsne}
\end{deluxetable*}

\begin{deluxetable*}{lccccccccc}
\small
\tablecolumns{10}
\caption{Karl G. Jansky Very Large Array Radio Observations at 6 GHz}
\tablehead{
\colhead{Source} & 
\colhead{RA} & 
\colhead{Dec} & 
\colhead{Observation Date} & 
\colhead{Bandpass Calibrator} &
\colhead{Phase Calibrator} & 
\colhead{$t_{\rm int}$} & 
\colhead{Beam Size} & 
\colhead{Beam Angle} & 
\colhead{rms} \\ 
\colhead{} & 
\colhead{[J200]} & 
\colhead{[J200]} & 
\colhead{} &
\colhead{} & 
\colhead{} & 
\colhead{[min]} & 
\colhead{[arcsec]} & 
\colhead{[deg]} & 
\colhead{[$\mu$Jy]}
}  
\startdata
SN1999as & 09:16:30.86 & +13:39:02.2 & 2017 Sep 16 & 3C48 & J0854+2006 & 37.80 & 1.30 $\times$ 1.03 & $-$59.42 & 6.67\\
SN2005ap & 13:01:14.84 & +27:43:31.4 & 2017 Nov 26 & 3C286 & J1310+3220 & 19.75 & 0.96 $\times$ 0.95 & $-$66.36 & 11.39\\
...&... &...&2019 Sep 03& 3C286 & J1310+3220 & 45.63 & 0.33 $\times$ 0.30 & 83.64 &4.69\\
SN2006oz & 22:08:53.56 & +00:53:50.4 & 2017 Sep 14 & 3C48 & J2212+0152 & 37.55 & 1.17 $\times$ 0.92 & 11.13 & 6.74\\
SN2007bi & 13:19:20.19 & +08:55:44.3 & 2017 Nov 26 & 3C286 & J1309+1154 & 19.75 & 1.18  $\times$ 1.04 & 85.43 & 12.14\\
SN2009jh & 14:49:10.08 & +29:25:11.4 & 2017 Dec 15 & 3C286 & J1443+2501 & 43.15 & 0.89 $\times$  0.86 & 22.65 & 6.50\\
PTF09cnd  & 16:12:08.94 & +51:29:16.1 & 2017 Dec 15 & 3C286 & J1549+5038 & 42.15 & 0.94 $\times$ 0.75 & 22.25 & 6.18\\
SN2010kd & 12:08:01.11 & +49:13:31.1 & 2017 Oct 26 & 3C286 & J1219+4829 & 37.55 & 1.09 $\times$ 0.88 & 76.33 & 4.78\\
SN2010gx & 11:25:46.71 & $-$08:49:41.4 & 2017 Oct 26 & 3C286 & J1131$-$0500 & 38.55& 1.88 $\times$ 0.90 & $-$42.57 & 5.64\\
SN2011ke  & 13:50:57.77 & +26:16:42.8 & 2017 Dec 02 & 3C286 & J1407+2827 & 41.15 & 1.05 $\times$ 0.90 & $-$54.68 & 4.97\\
SN2011kf & 14:36:57.53 & +16:30:56.6 & 2017 Dec 02 & 3C286 & J1446+1721 & 41.10 & 1.09 $\times$ 0.90 & $-$45.20 & 5.56\\
...&... &...&2019 Sep 03& 3C286 & J1446+1721 & 44.27 & 0.33 $\times$ 0.30 & 37.63 & 4.62\\
SN2011kg & 01:39:45.51 & +29:55:27.0 & 2017 Sep 15 & 3C48 & J0151+2744  & 40.80 & 1.36 $\times$ 0.96 & $-$79.72 & 6.30\\
SN2012il & 09:46:12.91 & +19:50:28.7 & 2017 Sep 16 & 3C48 & J0954+1743 & 40.5 &1.12 $\times$ 0.99 & $-$50.17 & 5.33 \\
...&... &... & 2019 Aug 27 & 3C286 & J0954+1743  & 45.60 & 0.31 $\times$ 0.27 & $-$1.32 & 5.32 \\
PTF12dam  & 14:24:46.20 & +46:13:48.3 & 2017 Dec 15 & 3C286 & J1417+4607 & 43.15 & 0.95 $\times$ 0.74 & 17.84 & 8.47\\
...&... &...&2019 Sep 03& 3C286 & J1417+4607 & 45.63 & 0.39 $\times$ 0.28 & $-$63.41 & 10.04\\
LSQ12dlf & 01:50:29.80 & $-$21:48:45.4 &  2017 Sep 15 & 3C48 & J0135$-$2008 & 40.80 & 1.15  $\times$ 1.04 & $-$32.01 & 8.96\\
\hline
GRB020903 & 22:48:42.34 & $-$20:46:09.3 & 2017 Sep 14 & 3C48 & J2243$-$2544 & 39.45 & 1.89 $\times$ 0.91 & 13.70 & 6.60\\
GRB030329  & 10:44:50.03 & +21:31:18.15 & 2017 Oct 24 & 3C48 & J1051+2119 & 40.50 & 1.48 $\times$ 1.04 & $-$70.47 & 5.60\\
GRB050826 & 05:51:01.59 & $-$02:38:35.4 & 2017 Sep 16 & 3C48 & J0541$-$0541 & 40.50 & 1.16 $\times$  0.92 & 0.96 & 6.46\\
GRB061021 & 09:40:36.12 & $-$21:57:05.4 & 2017 Oct 24 & 3C48 & J0927$-$2034 & 40.50 & 2.15 $\times$ 0.88 & $-$22.66 & 5.75\\
GRB090417B & 13:58:44.80 & +47:00:55.0 & 2017 Dec 02 & 3C286 & J1417+4607 & 41.20 & 0.96 $\times$ 0.89 & $-$25.99 & 5.67\\
GRB111225A & 00:52:37.22 & +51:34:19.5 &  2017 Sep 15 & 3C48 & J0105+4819 & 40.85 & 1.41 $\times$  0.87 & 84.14 & 6.00 \\
GRB120422A & 09:07:38.38 & +14:01:07.5  & 2017 Oct 24 & 3C48 & J0854+2006 & 38.80 & 1.30 $\times$ 1.05 & $-$62.28 & 5.07 \\
\enddata
\tablecomments{}
\label{tab:vla}
\end{deluxetable*}

\begin{deluxetable*}{lccccccccc}
\tablecolumns{10}
\caption{ALMA Millimeter Observations at 100 GHz}
\tablehead{
\colhead{Source} & 
\colhead{RA} & 
\colhead{Dec} & 
\colhead{Observation Date} & 
\colhead{Bandpass Calibrator} &
\colhead{Phase Calibrator} & 
\colhead{$t_{\rm int}$} & 
\colhead{Beam Size} & 
\colhead{Beam Angle} & 
\colhead{rms} \\ 
\colhead{} & 
\colhead{[J200]} & 
\colhead{[J200]} & 
\colhead{} &
\colhead{} & 
\colhead{} & 
\colhead{[min]} & 
\colhead{[arcsec]} & 
\colhead{[deg]} & 
\colhead{[$\mu$Jy]}
}  
\startdata
SN2006oz & 22:08:53.56 & +00:53:50.4 & 2017 Dec 17 & J2148+0657 & J2156-0037 & 21.67 & 0.41 $\times$ 0.31 & $-$85.96 & 17.61\\
SN2007bi & 13:19:20.19 & +08:55:44.3 & 2018 Mar 18 & J1229+0203 & J1254+1141 & 22.68 & 1.19 $\times$ 1.153 & $-$57.23 & 17.70\\
PTF10hgi & 16:37:47.00 & +06:12:32.3 & 2018 Jan 11 & J1550+0527 & J1658+0741 & 22.18 & 0.48 $\times$ 0.37 & 73.44 & 14.66\\
SN2010gx & 11:25:46.71 & $-$08:49:41.4 & 2018 Jan 12 & J1058+0133 & J1135$-$0428 & 21.17 &  0.53 $\times$ 0.39 & 61.30 & 19.04\\
SN2011kf & 14:36:57.53 & +16:30:56.6 & 2018 Jan 28 & J1550+0527 & J1446+1721 & 23.69 & 0.72 $\times$ 0.63 & 5.17 & 16.70\\
SN2012il & 09:46:12.91 & +19:50:28.7 & 2017 Dec 15 & J0854+2006 & J0940+2603 & 26.21 & 0.48 $\times$ 0.37 & 70.72 & 19.07\\
LSQ12dlf & 01:50:29.80 & $-$21:48:45.4 &  2017 Dec 30 & J0006$-$0623 & J0151$-$1732 & 20.66 & 0.40 $\times$ 0.29 & $-$76.78 & 17.06\\
SSS120810  & 23:18:01.80 & $-$56:09:25.6 & 2017 Dec 17 & J2357$-$5311 & J2336$-$5236 & 22.68 & 0.38 $\times$ 0.36 & $-$12.61 & 17.29\\
SN2013dg  & 13:18:41.35 & $-$07:04:43.0 & 2018 Jan 13 & J1256$-$0547 & J1312$-$0424 & 21.17 & 0.67 $\times$ 0.58 & 80.41 & 20.63\\
LSQ14bdq  & 10:01:41.60 & $-$12:22:13.4 & 2017 Dec 20 & J1037-2934 & J0957$-$1350 & 21.17 & 0.46 $\times$ 0.32 & 71.07 & 18.50\\
LSQ14mo  & 10:22:41.53 & $-$16:55:14.4 & 2017 Dec 20 & J1037$-$2934 & J0957$-$1350 & 20.66 & 0.47 $\times$ 0.32 & 72.08 & 20.22\\
LSQ14an  & 12:53:47.83 & $-$29:31:27.2 & 2017 Dec 04 & J1337-1257 & J1305$-$2850 & 20.66 & 0.25 $\times$ 0.21 & 87.49 & 16.11\\
LSQ14fxj  & 02:39:12.61 & +03:19:29.6 & 2017 Dec 29 & J0239+0416 & J0239+0416 & 22.18 & 0.42 $\times$ 0.30 & $-$76.52 & 22.18 \\
CSS140925  & 00:58:54.11 & +18:13:22.2 & 2018 Jan 01 & J0238+1636 & J0117+1418 & 25.71 & 0.54 $\times$ 0.32 & $-$52.06 & 16.68\\
SN2015bn  & 11:33:41.57 & +00:43:32.2 & 2018 Jan 12 & J1058+0133 & J1135$-$0428 & 21.67 & 0.57 $\times$ 0.39 & 55.43 & 18.22\\
OGLE15sd  & 01:42:21.46 & $-$71:47:15.6 & 2017 Dec 28 & J2357$-$5311 & J0112$-$6634 & 53.42& 0.45 $\times$ 0.36 & $-$22.87 & 12.08\\
SN2016ard  & 14:10:44.55 & $-$10:09:35.4 & 2018 Jan 25 & J1337$-$1257 & J1406$-$0848 & 21.17 & 0.82 $\times$ 0.58 & $-$86.36 & 16.93\\
iPTF16bad & 17:16:40 & +22:04:52.47 & 2019 Dec 12 & J1924$-$2914 & J1722+2815 & 69.55 &3.58 $\times$ 2.77 & $-$6.69 &  14.06\\
SN2016els & 20:30:13.92 & $-$10:57:01.81 & 2019 Oct 20 & J1924$-$2914 & J2025$-$0735 & 18.44 & 1.49 $\times$ 0.95 & 87.45 & 20.16\\
SN2017gci & 06:46:45.026 & $-$27:14:55.86& 2019 Dec 14 & J0538$-$4405 & J0632$-$2614 & 18.14 & 2.70 $\times$ 2.28 & 88.54 &  15.39\\
SN2017jan & 03:07:22.570 & $-$64:23:01.00 & 2019 Dec 25 & J0334$-$4008 & J0303$-$6211 & 21.17 & 3.31 $\times$ 2.68 & $-$42.71 & 21.46\\
SN2018bsz & 16:09:39.1 &	$-$32:03:45.73& 2019 Dec 17 & J1517$-$2422 &  J1556$-$3302 & 18.14 & 3.09 $\times$ 2.16 & $-$81.01 & 15.09\\
SN2018ffj & 02:30:59.77	& $-$17:20:26.82 & 2019 Dec 23 & J0006$-$0623 & J0212$-$1746 & 18.14 &3.31 $\times$ 2.34 & $-$89.71 &  17.81\\
SN2018ffs & 20:54:37.16 & +22:04:51.47 & 2019 Nov 11 & J2253+1608 & J2051+1743 & 29.23 & 1.96 $\times$ 1.66 & $-$8.35 & 17.99\\
SN2018gft & 23:57:17.95 & $-$15:37:53.27 & 2019 Dec 13 & J0006$-$0623 & J2354$-$1513 & 18.14 & 2.83 $\times$ 2.24 & 78.35 &  17.15\\
SN2018hti & 03:40:53.750 & +11:46:37.29 & 2019 Dec 26 & J0423$-$0120 & J0334+0800 & 20.16 & 2.89 $\times$ 2.66 & $-$41.55
 & 17.27\\
SN2018ibb & 04:38:56.960 &	$-$20:39:44.01 & 2019 Dec 24 & J0423$-$0120 & J0416$-$1851 & 18.14 & 3.03 $\times$ 2.39 & $-$71.73 & 19.19\\
SN2018jfo & 11:23:38.618 & +25:59:51.95 & 2019 Dec 31 & J1256$-$0547 &J1148+1840 & 44.86 & 3.82 $\times$ 2.39 & $-$36.95 & 17.93\\
SN2018lfe & 09:33:29.556 &	+00:03:08.39 & 2019 Dec 30 &J0750+1231 &J0930+0034 & 18.65 & 3.55 $\times$ 2.52 & 73.94 & 20.68\\
\hline
\enddata
\tablecomments{}
\label{tab:alma}
\end{deluxetable*}

\begin{deluxetable*}{lcccccccc}
\tablecolumns{9}
\caption{Source Parameters}
\tablehead{
\colhead{Source} & 
\colhead{$z$} & 
\colhead{Explosion Date} & 
\colhead{$\delta t_{\rm 6 \ GHz}^{\rm a}$} & 
\colhead{$\delta t_{\rm 100 \ GHz}^{\rm a}$} & 
\colhead{$F_{\rm 6 \ GHz}$} & 
\colhead{$F_{\rm 100\ GHz}$} & 
\colhead{$\rm SFR_{\rm radio}^{\rm b}$} & 
\colhead{$\rm SFR_{\rm opt}$} \\
\colhead{} & 
\colhead{} &
\colhead{[MJD]} & 
\colhead{[days]} & 
\colhead{[days]} & 
\colhead{[uJy]} & 
\colhead{[uJy]} & 
\colhead{[$\rm M_{\odot} \ yr^{-1}$]} & 
\colhead{[$\rm M_{\odot} \ yr^{-1}$]} 
}  
\startdata
SN1999as & 0.13 & 51172 & 6069 & -- & <20.01 & -- & <1.66 & 0.18\\
SN2005ap & 0.28 & 53424 & 3631, 4135 & -- & <34.17, <6.71  & -- & <16.31 & 0.13\\
SN2006oz & 0.38 & 54025 & 2896 & 2964 & <20.22 & <52.83 & <18.30 & 0.13\\
SN2007bi & 0.13 & 54075 & 3554 & 3653 & <36.42 & <53.10  &<3.07 & 0.02\\
SN2009jh & 0.35 & 55014 & 2288 & -- & <19.5 & -- & <15.00 & <0.01\\
PTF09cnd  & 0.26 & 55006 & 2461 & -- & <18.54 & -- & <7.19& 0.115\\
SN2010kd & 0.10 & 55499 & 2321 & -- & <14.34 & -- &<0.73 & 0.0\\
SN2010gx & 0.23 & 55251 & 2279 & 2341 & <16.92 & <57.12 & <5.09 & 0.26\\
SN2011ke  & 0.14 & 55651 & 2135 & -- & <14.91 & -- & <1.59 & 0.44\\
SN2011kf & 0.25 & 55921 & 1743, 2255 & 1787 & <16.68, <13.86 & <50.1 & <4.79 & 0.15\\
SN2011kg & 0.19 & 55912 & 1761 & -- & <18.9 & -- & <3.82 & 0.39\\
SN2012il & 0.18 & 55919 & 1782, 2386 & 1858 & <15.99, <15.96 & <57.21 & <2.64 & 0.40\\
PTF12dam  & 0.11 & 56022& 1878, 2445 & -- & 117.2 +/- 11.6, 61.5 $\pm 10.8^{\dagger}$ & -- & 6.75 & 4.8\\
LSQ12dlf & 0.26 & 56119& 1508 & 1592 & <29.76 & <51.18 & <11.24 & 0.40\\
SSS120810  & 0.16 & 56116 & -- & 1720 & -- & <51.87 & <55.12 & 0.06\\
SN2013dg  & 0.27 & 56419 & -- & 1353 & -- & <61.89 &  <10.15 & 0.4\\
LSQ14an  & 0.16 & 56513 & -- & 1357 &-- & <48.33 & <56.47 & 1.19 \\
LSQ14bdq  & 0.35 & 56721& -- & 1031 &  -- & <55.5 & <340.83 & 0.16\\
LSQ14fxj  & 0.36 & 56882& -- & 907 & -- & <53.34 & <360.76 & 0.74\\
LSQ14mo  & 0.25 & 56636 & -- & 1174 & -- & <60.66 & <185.77 & 0.52\\
CSS140925  & 0.46 & 56900 & -- & 835 & -- & <50.79 & <592.79 & 0.56\\
SN2015bn  & 0.11 & 57013& -- & 1003 & -- & <54.66 & <29.47 & 0.04\\
OGLE15sd  & 0.57 & 57295 & -- & 524 & -- & <36.24 & <690.66 & --\\ 
iPTF16bad & 0.24& 57513 & -- & 1056 & -- & <42.18& <122.14 & --\\
SN2016ard  & 0.20 & 57433 & -- & 590 & -- & <59.79 & <112.04 & 0.85\\
SN2016els & 0.22 & 57543 & -- & 1013 & -- & <60.48 & <131.91 & --\\
SN2017gci & 0.09& 57939& -- & 818 & -- & <46.17& <15.22 & -- \\
SN2017jan & 0.40& 57986 & -- & 613 & -- &<64.38 & <540.97& --\\
SN2018bsz & 0.03& 58203& -- & 615 & -- &<45.27 & <1.22 & 0.5\\
SN2018ffj & 0.23& 58275& -- & 458 & -- &<53.43 & <137.63 & --\\
SN2018ffs & 0.14& 58324& -- & 415 & -- & <53.97& <46.85& --\\
SN2018gft & 0.23& 58355& -- & 386 & -- &<51.45 & <127.58& --\\
SN2018hti & 0.06& 58415& -- & 403 & -- & <51.81& <8.12 & 0.39\\
SN2018ibb & 0.16& 58465& -- & 374 & -- & <57.57& <64.62 & --\\
SN2018jfo & 0.16 & 58411 & -- & 376 & -- & <53.79& <62.85& --\\
SN2018lfe & 0.35& 58424 & -- & 313 & -- & <62.04& <393.62 & 1.0\\
\hline
GRB020903 & 0.251 &  52884 & 4389 & -- & <19.80 & --& <7.22&2.65\\
GRB030329  & 0.1685 & 52727 &4557 &-- & <16.8 & -- &  <2.54 & 0.71\\
GRB050826 & 0.296 & 53608 & 3398 & --&  <25.83 & -- &  <10.21 & 1.17 \\
GRB061021 & 0.3463 & 54029& 2987&--&  <17.25 & -- &  <12.95 & 0.05\\
GRB090417B & 0.345 & 54938 & 2343&-- & <16.83 & -- &  <12.55 & 1.25\\
GRB111225A & 0.297 & 55920 & 1612&-- & <18.00 & -- & <9.56&--\\
GRB120422A & 0.283 & 56039 &1567 & --& <15.21 & -- &  <7.25& 1.65\\
\enddata
\tablecomments{Limits correspond to $3\sigma$. Explanations of explosion dates and references for $\rm SFR_{opt}$ given in Appendix \ref{appendix:a}.\\
$^{\rm a}$ Explosion rest-frame.\\
$^{\rm b}$ We use the most constraining limit for sources with multiple observations.\\
$^\dagger$ The host galaxy emission is resolved out in the second epoch, A-configuration observations.}
\label{tab:master}
\end{deluxetable*}

\begin{deluxetable*}{lcccccccc}
\tablecolumns{9}
\caption{SLSN Magnetar Parameters and CLOUDY Results}
\tablehead{
\colhead{Source} & 
\colhead{$P$} & 
\colhead{$B$} & 
\colhead{$M_{\rm ej}$} & 
\colhead{$E_K$} & 
\colhead{$v_{\rm ej}$} & 
\colhead{$t_{\rm ff, 6 GHz}$} & 
\colhead{$t_{\rm ff, 100 GHz}^{a}$} &
\colhead{$E_{\rm min,iso}^{b}$}\\ 
\colhead{} & 
\colhead{[ms]} & 
\colhead{[10$^{14}$ G]} &
\colhead{[$M_\odot$]} & 
\colhead{[$10^{51}$ erg]} & 
\colhead{[10$^3 \rm \ km \ s^{-1}$]} & 
\colhead{[yr]} & 
\colhead{[yr]} &
\colhead{[$10^{51}$ erg]}
}  
\startdata
SN2005ap & 1.28 & 1.71 & 3.57 & 8.85 & 15.22 & 2.43 & $<0.16$ &1108 \\
SN2006oz & 2.70 & 0.32 & 2.97 & 2.66 & 9.46 & 2.05 & $<0.19$ & 333 \\
SN2007bi & 3.92 & 0.35 &  3.80 & 2.37 & 7.90 & 5.94 & 1.94 & 297 \\
SN2009jh & 1.74 & 0.27 & 6.98 & 6.78 & 9.11 & 4.52 & 1.70 & 849\\
PTF09cnd & 1.46 & 0.10 & 5.16 & 3.29& 8.56 & 1.33 & $<0.16$  & 412\\
PTF10hgi & 4.78 & 2.03 & 2.19 & 0.55 & 5.12 & 4.85 & 1.39 & 69\\
SN2010gx & 3.66 & 0.59 & 2.39 & 3.78& 12.65 & 3.63 & 1.24 & 473\\
SN2010kd$^*$ & 3.51 & 0.57 & 10.51 & 3.64 & 5.90 & 16.13 & 4.43 & 456\\
SN2011ke & 0.78 & 3.88 & 7.64 & 5.22& 8.15 & 2.90 & $<0.13$  & 653\\
SN2011kf & 1.48 & 0.70 & 4.57 & 6.72 & 11.46 & 3.11 & $<0.15$  & 841\\
SN2011kg & 2.07 & 2.88 & 6.54 & 7.97 & 12.11 & 6.98 & 1.86 & 998\\
SN2012il & 2.35 & 2.24 & 3.14 & 1.94 & 7.93 & 3.95 & $<0.14$ & 243\\
PTF12dam & 2.28 & 0.18 & 6.27 &3.03 & 7.01 & 4.28 & 1.81 & 379\\
LSQ12dlf & 2.82 & 1.20 & 3.68 & 2.54& 8.28 & 5.72 & 1.66 & 318\\
SSS120810 & 3.00 & 1.93 & 2.22 &2.82 & 11.13 & 3.42 &  $<0.14$ & 353\\
SN2013dg & 3.50 & 1.56 & 2.75 & 1.85 &  8.38 & 5.19 & 1.50 & 232\\
LSQ14bdq & 0.98 & 0.49 & 33.71 & 25.06 & 8.71 & 16.67 & 4.57 & 3137\\
LSQ14mo & 4.97 & 1.01 & 2.10 & 2.43 &  10.74 &4.77 & 1.44 &304\\
SN2015bn & 2.16 & 0.31 & 11.73 &3.45 &  5.46 & 12.00 & 3.49 & 432\\
OGLE15sd & 2.16 & 1.74 & 1.91 &6.29 & 9.33 & 9.17 & 2.53 & 787\\
iPTF16bad & 3.73 & 2.62 & 2.22 & 1.12 & 7.11 & 4.55 & 1.29 & 140\\
SN2016ard$^{\dagger}$ &  0.93 & 1.55 & 16.6 &33.0 & 14.2 & 7.80 & <2.11 & 4131\\
SN2016els$^*$ & 0.92 & 5.38 & 11.83 & 28.0 & 15.43 & 5.50 & 1.39 &  3505\\
SN2017gci$^*$ & 1.26 & 3.46 & 11.74 & 7.16 & 7.83 & 6.86 & 1.81 & 896\\
SN2017jan$^*$ & 3.08 & 0.34 & 7.14 & 4.58 & 8.03 & 10.84 & 3.26 & 573 \\
SN2018hti$^*$ & 1.25 & 2.59 & 31.04 &11.3 & 6.06 & 15.67 & 3.80 & 1415\\
SN2018ibb$^*$ & 0.74 & 0.16 & 43.47 & 45.6 & 10.27 & 12.12 & 3.67 & 5708 \\
SN2018lfe$^{\ddagger}$ & 2.85 & 2.2 & 3.80 & 3.82& 10.05 & 6.04 & 1.71 & 478\\
\enddata
\tablecomments{Magnetar parameters are from \citet{Nicholl2017b} except where specified. All times refer to the observer frame.\\$^{\rm a}$ We report upper limits where $t_{\rm ff,100GHz} < 1 \rm yr$ $(1+z)$, as these results are based on extrapolations to early times.\\$^{\rm b}$ Minimum energy required for a $10^\circ$ jet to break out of the SN ejecta. \\$^{*}$ Parameters derived in this work.  \\${\dagger}$ From \citet{Blanchard2018}.\\${\ddagger}$ From Yin et al. \textit{in prep.}}
\label{tab:magnetars}
\end{deluxetable*}

\clearpage
\newpage
\appendix
\section{Brief Description of Sources in our Sample}
\label{appendix:a}
\subsection{SN1999as}
SN1999as ($z \approx 0.127$) was discovered on 1999 February 18 by the Supernova Cosmology Project (SCP) \citep{Knop1999}. From \citet{Knop1999}, the peak time is MJD 51242. Given its similarity to SN2007bi, we assume a 62 d rest-frame rise time, corresponding to an explosion date of MJD 51172. Spectroscopic data are given in \citet{Hatano2001}. Observations of the host galaxy are given in \citet{Leloudas2015}, \citet{Angus2016}, and \citet{Schulze2018}.

\subsection{SN2005ap}
SN2005ap ($z \approx 0.28$) was discovered on 2005 March 3 in images taken with the Robotic Optical Transient Search Experiment Telescope (ROTSE-IIIb; \citealt{Akerlof2003}) as part of the Texas Supernova Search (TSS; \citealt{Quimby2006}). SN2005ap exhibited a 1-3 week rise to peak and a relatively rapid decay \citep{Quimby2007}. We assume a rest-frame rise time of 12 days and a peak time of MJD 53439 \citep{Quimby2011}, corresponding to an explosion date of 53424. Radio and optical observations of the host are presented in \citep{Schulze2018}. Additional host galaxy observations are given in \citet{Lunnan2014}.

\subsection{SN2006oz}
SN2006oz ($z \approx 0.38$) was discovered on 2006 October 20 by the Sloan Digital Sky Survey II. We assume an explosion date of MJD 54025. Light curves and spectroscopy are given in \citet{Leloudas2012}. Host galaxy observations are given in \citet{Lunnan2014}, \citet{Leloudas2015}, and \citet{Schulze2018}.

\subsection{SN2007bi}
SN2007bi ($z \approx 0.13$) was discovered on 2007 April 6 by the Nearby Supernova Factory \citep{Nugent2007}. The SN exhibited a slow rise time of 77 days with a peak time of MJD 54152, corresponding to an explosion date of MJD 54075 \citep{Gal-Yam2009}. Host galaxy observations are given in \citet{Lunnan2014} and \citet{Schulze2018}. The relatively slow decay time and the large inferred mass of radioactive $^{56}$Ni have been used to argue for a pair-instability supernova explosion \citep{Gal-Yam2009}, although \citet{Nicholl2013} have shown that the event properties are well-matched to a magnetar central engine with a modest ejecta mass. 

\subsection{SN2009jh}
SN2009jh (= PTF09cwl = CSS090802:144910+292510; $z \approx 0.35$) was first discovered in the Catalina Sky Survey on 2009 August 2 \citep{Drake2009} and independently during commissioning of the PTF system \citep{Quimby2011}. From \citet{Quimby2011}, the peak time is MJD 55081 and the rest-frame rise time is $\sim 50$ days, corresponding to an explosion date of 55014. Light curves and spectroscopy are given in \citet{DeCia2018} and \citet{Quimby2018}. Observations of the host galaxy are presented in \citet{Leloudas2015}, \citet{Angus2016}, \citet{Perley2016a} and \citet{Schulze2018}. Extensive X-ray observations with \textit{Swift-}XRT spanning $\delta t = 48-1961$ d reveal no X-ray source at the location of the SN (\citealt{Levan2013}; \citealt{Margutti2018}).

\subsection{PTF09cnd}
PTF09cnd ($z \approx 0.26$) was first detected on 2009 July 13 by the Palomar Transient Factory with the 1.2 m Samuel Oschin Telescope during commissioning of the PTF system \citep{Quimby2011}. From \citet{Inserra2013}, the peak time is MJD 55069.145 and the rest-frame rise-time is $\sim 50$ days. Thus the explosion date is MJD 55006. Spectra for PTF09cnd are given in \citet{Quimby2018}. Observations and properties of the host galaxy are presented in \citet{Neill2011}, \citet{Leloudas2015}, and \citet{Perley2016a}.  A previous search for FRBs from this event was conducted in \citet{Hilmarsson2020}. X-ray non-detections place limits on the unabsorbed fluxes between $10^{-13} - 10^{-15} \rm \ erg \ cm^{-2} \ s^{-1}$ (\citealt{Levan2013}; \citealt{Margutti2018}).

\subsection{SN2010kd}

SN2010kd ($z \approx 0.23$) was discovered on 2010 November 14 by the Robotic Optical Transient Search Experiment Telescope (ROTSE-IIIb; \citealt{Akerlof2003}). From \citet{Vinko2012}, we assume a peak time of MJD 55554 and a rest-frame rise time of 50 days, corresponding to an explosion date of MJD 55499. Host galaxy observations are given in \citet{Leloudas2015} and \citep{Schulze2018}. X-ray observations of the source span $\delta t \sim 120-1964$ days post-explosion (rest-frame) and correspond to limits on the unabsorbed X-ray flux of $10^{-13} - 10^{15} \rm \ erg \ cm^{-2} \ s^{-1}$ \citep{Levan2013,Margutti2018}.

\subsection{SN2010gx}
SN2010gx (= PTF10cwr = CSS100313:112547-084941; $z \approx 0.23$) was discovered by the Catalina Real-time Transient Survey on 2010 March 13 \citep{Mahabal2010}. From \citet{Inserra2013}, the peak time is MJD 55279 and the rest-frame rise time is $\sim 23$ days; the explosion date is therefore MJD 55251. Light curves and spectra are given in \citet{Quimby2011} and \citet{Inserra2013}. Host galaxy observations are given in \citet{Chen2013}, \citet{Lunnan2014}, \citet{Leloudas2015}, \citet{Perley2016a} and \citet{Schulze2018}. A previous search for FRBs from this event was conducted in \citet{Hilmarsson2020}. \textit{Swift}-XRT observations spanning $\delta t = 19 - 659$ days reveal no X-ray source at the location of the SN \citep{Levan2013,Margutti2018}.

\subsection{SN2011ke}

SN2011ke (= PTF11dij = CSS110406:135058+261642 = PS1-11xk; $z \approx 0.14$) was discovered by the Catalina Real-time Transient Survey on 2011 April 4 \citep{Drake2011} and independently by the Palomar Transient Factory on 2011 March 30 \citep{Quimby2011_sn2011ke}. A non-detection of the event one day prior to the 2011 March 30 detection constrains the explosion date to MJD 55650.65 \citep{Inserra2013}. Host galaxy observations are given in \citet{Lunnan2014}, \citet{Leloudas2015}, \citet{Perley2016a}, and \citet{Schulze2018}. \textit{Swift}-XRT observations span $\delta t \sim 40 - 1604$ days post-explosion and reveal no X-ray source at the transient position \citep{Levan2013,Margutti2018}.

\subsection{SN2011kf}

SN2011kf (= CSS111230:143658+163057; $z \approx 0.25$) was discovered on 2011 December 30 by the Catalina Real-tme Transient Survey \citep{Drake2012}. Additional spectra were obtained by \citet{Prieto2012}. From \citet{Inserra2013}, the inferred explosion date is MJD 55920.65. Host galaxy observations are given in \citet{Lunnan2014}, \citet{Leloudas2015} and \citep{Schulze2018}. 

\subsection{SN2011kg}
SN2011kg (= PTF11rks; $z \approx 0.19$) was discovered on 2011 December 21 by the Palomar Transient Factory \citep{Quimby2011_sn2011kg}. From \citet{Inserra2013}, the explosion date is MJD 55912.1. Host galaxy photometry and spectroscopy are given in \citet{Lunnan2014} and \citet{Perley2016a}. \textit{Swift}-XRT observations reveal no X-ray source at the position of the SN over the time period $\delta t \sim 11 - 25$ days \citep{Levan2013,Margutti2018}.

\subsection{SN2012il}
SN2012il (= PS1-12fo = CSS120121:094613+195028; $z \approx 0.18$) was discovered by the Pan-STARRS1 3Pi Faint Galaxy Supernova Survey on 2012 January 19 \citep{Smartt2012} and independently by the Catalina Real-time Transient Survey on 2012 January 21 \citep{Drake2012}. From \citet{Inserra2013}, the explosion date is MJD 55918.56. Observations of the host galaxy are presented in \citet{Lunnan2014}, \citet{Leloudas2015} and \citet{Schulze2018}. No X-ray emission is detected at the location of the source down to $\sim 10^{-16} \ \rm erg \ s^{-1} \ cm^{-2}$ \citep{Levan2013,Margutti2018}.

\subsection{PTF12dam}
PTF12dam ($z \approx 0.11$) was discovered by the Palomar Transient Factory on 2012 April 10 \citep{Quimby2012}, and is among the subset of slowly evolving SLSNe (\citealt{Nicholl2013}; \citealt{Inserra2017}; \citealt{Vreeswijk2017}; \citealt{Quimby2018}). From \citet{Nicholl2013}, the peak date is MJD 56088, and the rest-frame rise time is $\sim 60$ days; the inferred explosion date is thus MJD 56022. Extensive studies of the host galaxy are presented in the literature \citep{Lunnan2014,Chen2015,Leloudas2015,Thone2015,Perley2016a,Hatsukade2018}. The source is not detected in $\textit{Swift}$-XRT observations over the timescale $\delta t \sim 43-900$ days down to a limiting flux of $F_X \sim 5 \times 10^{-14} \ \rm erg \ s^{-1} \ cm^{-2}$ \citep{Margutti2018}. \textit{Chandra} X-ray observations over the range $\delta t \sim 60 - 68$ d reveal an X-ray source with an unabsorbed flux $F_x = (7.3 \pm 2.9) \times 10^{-16}  \ \rm erg \ s^{-1} \ cm^{-2}$ (0.3$-$10 keV) \citep{Margutti2018}. A previous search for FRBs from this event was conducted in \citet{Hilmarsson2020}. 
 
\subsection{LSQ12dlf} 
LSQ12dlf ($z \approx 0.26$) was discovered by the La Silla-Quest survey (LSQ; \citealt{Baltay2013}) on 2010 July 10 and subsequently classified as a SLSN by the Public ESO Spectroscopic Survey of Transient Objects (PESSTO) \citep{Smartt2012_atel}. From \citet{Nicholl2014}, the peak time is MJD 56149 and the rest-frame Rise time is approximately $\sim 24$ days. The inferred explosion date is therefore MJD 56119. A previous search for FRBs from this event was conducted in \citet{Hilmarsson2020}. Host galaxy observations are given in \citep{Schulze2018}. 

\subsection{SSS120810}
SSS120810 ($z \approx 0.16$) was discovered on 2012 August 10 by the Catalina Real-time Transient Survey and subsequently classified as a SLSN by PESSTO \citep{Wright2012}. From \citet{Nicholl2014}, the peak time is MJD 56146. Assuming a rest-frame rise time of 26 days, the explosion date is MJD 56116. Observations of the host galaxy are given in \citet{Leloudas2015} and \citet{Schulze2018}.

\subsection{SN2013dg}
SN2013dg (= CSS130530:131841-070443 = MLS130517:131841-07044; $z \approx 0.27$) was discovered on 2013 May 17 by the Mount Lemmon Survey and independently by the Catalina Sky Survey on 2018 May 30 \citep{Drake2013}. From \citet{Nicholl2014}, the peak time is MJD 56449 and the rest-frame rise time is approximately 24 days. The Inferred explosion date is therefore MJD 56419. Host galaxy observations are given in \citet{Schulze2018}.

\subsection{LSQ14bdq}
LSQ14bdq ($z \approx 0.35$) was discovered by the LSQ on 2014 April 5 and subsequently classified by PESSTO \citep{Benitez2014}. The lightcurve exhibits an initial peak lasting $\sim 15$ days followed by a slower rise to maximum light \citep{Nicholl2015}. Non-detections prior to the initial peak constrain the explosion date to MJD 56721. Host galaxy observations are given in \citet{Schulze2018}. 

\subsection{LSQ14mo}
LSQ14mo ($z \approx 0.25$) was discovered by the LSQ on 2014 January 30 and subsequently classified by PESSTO as a SLSN on 2014 January 31 \citep{Leloudas2014_atel}. From \citet{Leloudas2015}, the peak time is MJD 56699. Assuming a rest-frame rise time of 50 days, the inferred explosion date is MJD 56636. Host galaxy observations are given in \citet{Chen2017} and \citet{Schulze2018}. \textit{Swift}-XRT observations over the timescale $\delta t \sim 52 - 774$ days rest-frame post explosion do not show evidence for X-ray emission \citep{Margutti2018}.

\subsection{LSQ14an}
LSQ14an ($z \approx 0.16$) was discovered by the LSQ on 2014 January 3 and classified as a SLSN by PESSTO \citep{Leget2014}. The light curve indicates LSQ14an is part of the class of slowly evolving SLSN \citep{Inserra2017}.  From \citet{Jerkstrand2017}, the peak time is MJD 56595. Assuming a rest-frame rise time of $\sim 70$ days, the inferred explosion date is MJD 56513. Host galaxy observations are given in \citep{Schulze2018}. No X-ray emission is detected at the location of the source based on \textit{Swift}-XRT observations spanning $\delta t \sim  64 - 234$ days rest-frame post explosion \citep{Margutti2018}

\subsection{LSQ14fxj}
LSQ14fxj ($z \approx 0.36$) was discovered by the LSQ on 2014 October 12 and classified as a SLSN I by PESSTO \citep{Smith2014}. From \citet{Smith2014}, on MJD 56940, the source was 4$-$5 weeks post-maximum light. We assume a rest-frame rise time of 50 days, as in \citet{Margutti2018}. The explosion date is MJD 56882. Host galaxy observations are given in \citep{Schulze2018}. \textit{Swift}-XRT observations of the source span $\delta t \sim 64 - 234$ days rest frame post-explosion, and do not reveal any X-ray emission at the location of the source \citep{Inserra2017,Margutti2018}.

\subsection{CSS140925}
CSS140925 ($z \approx 0.46$) was discovered on 2014 September 25 by the Catalina Sky Survey and classified by PESSTO  \citep{Campbell2014}. From the CRTS catalog (http://nesssi.cacr.caltech.edu/catalina/AllSN.html), the explosion date is MJD 56900. No X-ray emission is detected at the SN location over the timescale $\delta t \sim 29 - 186$ days rest-frame post-explosion \citep{Margutti2018}. Host galaxy observations are given in \citep{Schulze2018}

\subsection{SN2015bn}
SN2015bn (= PS15ae = CSS141223-113342+004332 = MLS150211-113342
+004333; $z \approx 0.11$) was initially discovered by the Catalina Sky Survey on 2014 December 23 \citep{Drake2009}. It was subsequently discovered by the Mount Lemmon Survey on 2015 February 11 and the PanSTARRS Survey for Transients on 2015 February 15 \citep{Huber2015}. Host galaxy observations are presented in \citet{Schulze2018}. Spectroscopy and UV to NIR photometry spanning -50 to +250 days from optical maximum revealed a slowly fading source with undulations in the light curve on a timescale of 30 $-$ 50 days \citep{Nicholl2016a}. Late-time, nebular-phase observations show evidence for a central engine (\citealt{Nicholl2016b}; \citealt{Jerkstrand2017}). Radio emission is not detected at 320 $-$ 335 days post explosion \citep{Nicholl2016a}. X-ray observations correspond to upper limits on the unabsorbed flux in the range  $10^{-14} - 10^{-13} \rm \ erg \ cm^{-2} \ s^{-1}$ \citep{Margutti2018}.

\subsection{OGLE15sd}
OGLE15sd ($z \approx 0.57$) was discovered on 2015 November 7 by the Optical Gravitational Lensing Experiment (OGLE)  \citep{Wyrz2015}. From the OGLE-IV Transient Detection System, the explosion date is MJD 57295 (http://ogle.astrouw.edu.pl/ogle4/transients/transients.html). No X-ray emission is detected at the SN location over the timescale $\delta t \sim 34 - 212$ days rest-frame post-explosion \citep{Margutti2018}.

\subsection{iPTF16bad} 
iPTF16bad ($z \approx 0.2467$) was discovered on 2016 May 31 by the Intermediate Palomar Transient Factory (iPTF) \citep{Yan2017}. The event is among a small subset of SLSNe displaying H$\alpha$ emission at late times. We constrain the explosion date to MJD
57513 by fitting the optical light curve to a magnetar model using the Modular Open-Source Fitter for Transients (\texttt{MOSFit}).

\subsection{SN2016ard}
SN2016ard (= PS16aqv = CSS160216:141045-100935; $z \approx 0.16$) was discovered on 2016 February 20 by the  Pan-STARRS near-Earth object survey \citep{Chambers2016} and classified by \citet{Chornock2016}. From \citet{Blanchard2018}, the peak time is MJD 57453 and the rest-frame rise time is 17 days, corresponding to an explosion date of MJD 57433.4. \textit{Swift}-XRT observations span $\delta t \sim 53 - 131$ days rest-frame since explosion and do not show evidence for X-ray emission at the SN location \citep{Margutti2018}.

\subsection{SN2016els}
SN2016els (= AT2016els = PS16dnq; $z \approx 0.217$) was discovered on 2016 July 29 by PESSTO \citep{Mattila2016}. From the Open Supernova Catalog, the peak date is MJD 57604. We assume a 50 day rest-frame rise time for an explosion date of MJD 57543.

\subsection{SN2017gci}
SN2017gci (= AT2017gci = Gaia17cbp; $z \approx 0.09$) was discovered on 2017 August 12 by the Gaia Photometric Survey \citep{Delgado2017} and subsequently classified as a SLSN by ePESSTO \citep{Lyman2017}. We constrain the explosion date to MJD 57939 by fitting the optical light curve to a magnetar model using \texttt{MOSFit}.

\subsection{SN2017jan}
SN2017jan (= AT2017jan = OGLE17jan; $z \approx 0.396$) was discovered on 2017 December 18 by the Optical Gravitational Lensing Experiment (OGLE) \citep{Wyr2017} and subsequently classified by ePESSTO \citep{Angus2017}. We constrain the explosion date to MJD 57986 by fitting the optical light curve to a magnetar model using \texttt{MOSFit}.

\subsection{SN2018bsz}
SN2018bsz (= ASASSN-18km = AT2018bsz = ATLAS18pny; $z \approx 0.0267$) was discovered on 2018 May 17 by the All Sky Automated Survey for SuperNovae (ASAS-SN) \citep{Brimacombe2018} and detected independently by the Asteroid Terrestrial-impact Last Alert System (ATLAS) on 2018 May 21. From \citet{Anderson2018}, the explosion date is estimated as the mid-point between the last non-detection epoch and the discovery date and is given by MJD 58202.5. Properties of the host galaxy are also given in \citet{Anderson2018}.

\subsection{SN2018gft}
SN2018gft (= AT2018gf = ZTF18abshezu = ATLAS18uymActions; $z \approx 0.23$) was dicovered on 2018 September 2 by ZTF \citep{Fremling2018_sn2018gft}. We constrain the explosion date to MJD 58355 by fitting the optical light curve to a magnetar model using \texttt{MOSFit}.

\subsection{SN2018ffj}
SN2018ffj (= AT2018ffj = ATLAS18tec; $z \approx 0.234$) was discovered on 2018 August 7 by ATLAS \citep{Tonry2018_sn2018ffj} and classified as a SLSN by ePESSTO on 2018 August 19 \citep{Kos2018}. From the Open Supernova Catalog, the peak date is MJD 58337. We assume a 50 day rest-frame rise time, corresponding to an explosion date of MJD 58275.

\subsection{SN2018ffs}
SN2018ffs (= AT2018ffs = ATLAS18txu = ZTF18ablwafp; $z \approx 0.142$) was discovered on 2018 August 13 by ATLAS \citep{Tonry2018_sn2018ffj} and classified by ePESSTO on 2018 September 16 \citep{Gromadzki2018}. We constrain the explosion date to MJD 58324 by fitting the optical light curve to a magnetar model using \texttt{MOSFit}.

\subsection{SN2018hti}
SN2018hti (= AT2018hti = ATLAS18yff = MLS181110:034054+114637 = Gaia19amt; $z \approx 0.063$) was discovered on 2018 November 2 by ATLAS \citep{Tonry2018_sn2018hti} and classified by the Global Supernova Project \citep{Burke2018}. We constrain the explosion date to MJD 58415 by fitting the optical light curve to a magnetar model using \texttt{MOSFit}.

\subsection{SN2018ibb}
SN2018ibb (= AT2018ibb = ATLAS18unu = ZTF18acenqto = Gaia19cvo; $z \approx 0.16$) was discovered on 2018 September 10 by ATLAS \citep{Tonry2018} and subsequently classified by ZTF \citep{Freling2018_sn2018ibb}. From the Open Supernova Catalog, the peak date is MJD 58465. We assume a $50$ day rest-frame rise time, corresponding to an explosion date of MJD 58407.

\subsection{SN2018jfo}
SN2018jfo (= AT2018jfo = ZTF18achdidy = MLS181220:112339+255952; $z \approx 0.163$) was discovered on 2018 November 10 by the Zwicky Transient Facility \citep{Fremling2018} and subsequently classified as a SLSN on 2019 January 5 \citep{Fremling2019}. We constrain the explosion date to MJD 58411 by fitting the optical light curve to a magnetar model using \texttt{MOSFit}.

\subsection{SN2018lfe}
SN2018lfe (= AT2018lfe = PS18cpp = ZTF18acqyvag; $z \approx 0.35$) was discovered on 2018 December 31 by the Pan-STARRS Survey for Transients \citep{Chambers2019} and subsequently classified as a SLSN on 2019 February 6 \citep{Gomez2019}. 
We constrain the explosion date to MJD 58424 by fitting the optical light curve to a magnetar model using \texttt{MOSFit}.

\subsection{GRB 020903}
GRB 020903 (=  XRF020903; $z\approx 0.251$) was discovered on 2002 September 3 by the Wide-Field X-Ray monitor and the Soft X-Ray Camera on the High Energy Transient Explorer-2 (HETE-II) \citep{Soderberg2004}. Observations of the radio and optical afterglow are presented in \citet{Soderberg2004}. Host galaxy observations are given in \citet{Michaowski2012} and \citet{Greiner2016}.

\subsection{GRB 030329}
GRB 030329 ($z \approx 0.168$) was discovered by the HETE-II satellite 2003 March 29 \citep{Vanderspek2003}. Radio observations of the afterglow, which is observable to $\delta t \sim 5$ yr post-explosion, are presented in \citet{Mesler2012} and \citet{vanderhorst2008}. Optical and radio observations of the host galaxy are presented in \citet{Niino2017} and \citet{Michaowski2012}, respectively.

\subsection{GRB 050826}
GRB 050826 ($z \approx 0.296$) was discovered on 2005 August 26 by the \textit{Swift} Burst Alert Telescope (BAT) \citep{Mangano2005}. The event is categorized as a sub-luminous, sub-energetic event \citep{Mirabal2007}. A previous search for FRBs from this event was conducted in \citet{Hilmarsson2020}. Host galaxy observations are presented in \citet{Levesque2010} and \citet{Niino2017}. 

\subsection{GRB 061021}
GRB 061021 ($z \approx 0.3463$) was discovered by the \textit{Swift}-BAT on 2006 October 21 \citep{Moretti2006}. The event is one of a small subset of GRBs that exhibit excess X-ray emission at early times, which has been attributed to thermal emission from the shock breakout of a supernova \citep{Sparre2012}. Optical and radio observations of the host galaxy are presented in \citet{Michaowski2012,Perley2015,Greiner2016}.

\subsection{GRB 090417B}
GRB 090417B ($z \approx 0.345$) was discovered on 2009 April 17 by the \textit{Swift}-BAT \citep{Sbarufatti2009}. The burst is classified as a ``dark'' GRB, a small subset of events lacking optical afterglows. Optical and radio observations of the host galaxy are given in \citet{Perley2013a} and \citet{Perley2013b}, respectively.

\subsection{GRB 111225A}
GRB 112225A ($z \approx 0.297$) was discovered on 2011 December 25 by the \textit{Swift}-BAT \citep{Siegel2011}. A previous search for FRBs from this event was conducted in \citet{Hilmarsson2020}. Host galaxy observations are not reported in the literature for this event.

\subsection{GRB 120422A}
GRB 120422A ($z \approx 0.283$) was discovered and localized by the \textit{Swift}-BAT on 2012 April 22 \citep{Troja2012}. The event is classified as a transition object between low- and high-luminosity GRBs \citep{Schulze2014}. Host galaxy observations are presented in \citet{Niino2017}.

\bibliographystyle{aasjournal}
\bibliography{ref}

\end{document}